\documentclass[eqsecnum,twocolumn,aps,tightenlines,floats,prd,showpacs]{revtex4-1}
\def\bea{\begin{eqnarray}}
\def\eea{\end{eqnarray}}
\def\bal{\begin{align}}
\def\eal{\end{align}}

\def\sfrac#1#2{{\textstyle \frac{#1}{#2}}}

\usepackage{graphics}
\usepackage{graphicx}
\usepackage{epsf} 
\usepackage{amsmath}
\usepackage{amssymb}
\usepackage{bbold}
\usepackage{slashed}
\usepackage{appendix}
\usepackage{bm}

\begin{document}  


\phantom{0}
\hspace{5.5in}\parbox{1.5in}{ \leftline{JLAB-THY-14-1863}
                \leftline{}\leftline{}\leftline{}\leftline{}
}
\title
{\bf  Covariant Spectator Theory of $np$ scattering: \\Isoscalar interaction currents}

\author{Franz Gross$^{1,2}$}
\email{email address: gross@jlab.org}

\affiliation{
$^1$Thomas Jefferson National Accelerator Facility, Newport News, VA 23606 \vspace{-0.15in}}
\affiliation{
$^2$College of William and Mary, Williamsburg, Virginia 23185}


\date{\today}

\begin{abstract} 


Using the Covariant Spectator Theory (CST),  one boson exchange (OBE) models have been found that give precision fits to low energy $np$ scattering and the deuteron binding energy. The  boson-nucleon vertices used in these models contain a momentum dependence that requires a new class of interaction currents for use with electromagnetic interactions.  Current conservation requires that these new interaction currents satisfy a two-body Ward-Takahashi identity, and using  principals of {\it simplicity\/} and {\it picture independence\/}, these currents can be uniquely determined. The results lead to general formulae for a  two-body current that can be expressed in terms of relativistic $np$ wave functions, ${\it \Psi}$, and two convenient truncated wave functions, ${\it \Psi}^{(2)}$ and $\widehat {\it \Psi}$, which contain all of the information needed for the explicit evaluation of the contributions from the interaction current.   These three wave functions can be calculated from the CST bound or scattering state equations (and their off-shell extrapolations).  A companion paper uses this formalism to evaluate the deuteron magnetic moment.


\end{abstract}
 
\phantom{0}
\vspace{0.76in}
\vspace*{-0.1in}  

\maketitle


\section{Introduction and Summary}

\subsection{Background}

Very quickly after the discovery of the both the deuteron and the neutron in 1932, the measured deuteron magnetic moment was used to estimate the neutron magnetic moment (for an early review see Ref.\  \cite{Bethe:1936zz}).  Calculations of the deuteron electromagnetic form factors, along with the magnetic and quadrupole moments, have long been a definitive test of hadronic theory.  The first  calculations the deuteron form factors for large momentum transfer  \cite{Jankus1956} and the first measurement  \cite{McIntyre1956} were published in 1956.    For recent reviews see Refs.~\cite{Garcon:2001sz,Gilman:2001yh}.

This work is the first of a series of  four planned papers (the second paper, referred to as Ref.\ II \cite{RefII}, accompanies this paper) that present the fourth generation calculation of the deuteron form factors  using what is now called the Covariant Spectator Theory (CST).   Like all work on this subject, study of the form factors has proved to be an important doorway into the development of hadronic theory.

I did the first of these calculations in 1964-65  using dispersion theory \cite{Gross:1964mla,Gross:1964zz}, where the form factors are expressed as dispersion integrals in the square of the momentum transferred by the virtual photon, $q^2=-Q^2$.  From this point of view the virtual photon couples to the $d\bar d$ channel, and the role of the nucleon appears through contributions from the $N\bar N$ intermediate states.   The normal threshold for the $N\bar N$ cut is at $4m^2$, way above the three-pion threshold at $9m_\pi^2$ (the deuteron is isoscalar and $G$-parity conservation prevents if from coupling to the lower threshold two-pion intermediate state).  The existence of anomalous thresholds explains this paradox: the diagram with a nucleon exchanged between the $d\bar d$ pair contributes a  singularity with an ``anomalous'' threshold at only $16mE_B\simeq 1.7 m_\pi^2$ (with $E_B=2.2246$ MeV the deuteron binding energy), significantly below the three-pion cut, establishing, even from this novel point of view, the overwhelming importance of the $d\to NN$ (or, simply, $dNN$) vertex function to our understanding of the electromagnetic structure of the deuteron \cite{Gross:1966zz}.  Perhaps more significantly, the imaginary parts of the form factors in the anomalous region come from contributions where the exchange nucleon is {\it on mass-shell\/}.  This exchange nucleon plays the role of a {\it spectator\/}, removed from the interaction of the virtual photon with the $N \bar N$ pair.  At the same time it was realized that  many other diagrams, with pions dressing the $dNN$ vertices,  also produce anomalous thresholds, and that these should  be summed to all orders in order to naturally regularize the leading anomalous contributions.  From this observation it was a short step to the idea that the description is unified \cite{Gross:1965zz} by introducing a covariant wave function satisfying an integral equation in which one nucleon is on-shell (as dictated by the anomalous cut condition), leading naturally to the CST equation for the $dNN$ vertex.   Study of the deuteron form factor lead directly to the introduction of the CST.  Only later \cite{Gross:1969rv,Gross:1993zj}  was it realized that the {\it cancellation theorem\/} provided another, perhaps more convincing, justification for the CST.

The second CST calculation of the deuteron from factors was done in 1980 in collaboration with  Arnold and Carlson \cite{SLAC-PUB-2318}.  That paper evaluated twice the  contribution from diagram (A) in Fig.~\ref{Fig1},  now described as the  Relativistic Impulse Approximation (RIA).  At that time we did not have relativistic deuteron wave functions determined directly from the $np$ scattering data; we used  wave functions either  constructed from nonrelativistic models or the Buck-Gross wave functions \cite{Buck:1979ff} constrained only by the deuteron binding energy and quadrupole moment.  

The third generation calculations, done in 1995 in collaboration with  Van Orden and Devine \cite{VanOrden:1995eg},  used, for the first time, consistent covariant wave functions determined by fitting the $np$ phase shifts up to lab energies of 350 MeV \cite{Gross:1991pm}.  At that time we realized that a proper gauge invariant calculation \cite{Gross:1987bu} required all four of the diagrams shown in Fig.~\ref{Fig1}, and results for both the RIA and the Complete Impulse Approximation (CIA) were presented.  This calculation was successful, showing that a fully relativistic calculation could provide a very good explanation of the form factors, even up to high $Q^2$.  

At the same time that  the third generation calculations were being done, Alfred Stadler and I were working on extending the CST formalism to the three-nucleon sector.  Using the CST  three body equations \cite{Stadler:1997iu} we discovered in 1996 \cite{Stadler:1996ut} that a reasonable description of the triton binding energy could not be obtained unless  the one boson exchange (OBE) models used previously were extended by  adding off-shell couplings to the $NNs$ vertices, $\Lambda^s$, for  the exchange of the scalar meson $s$.  Written in terms of the   off-shell projection operator 
\bea
\Theta(p)=\frac{m-\slashed{p}}{2m} \label{eq:theta}
\eea
the $NNs$ vertices now become 
\bea
\Lambda^s(p,p')=g_s{\bf 1}\to g_s{ \bf 1}-\nu_s\big[\Theta(p)+\Theta(p')\big]\qquad
\label{eq:1.2}
\eea
where $\nu_s$ is a new parameter to be determined by fitting data, and $p$ and $p'$ are the four-momenta of the outgoing and incoming nucleons, respectively.  It is important to realize that these factors are non-zero whenever particles are off-shell, and hence are a feature of Bethe-Salpeter or CST equations.   We  discovered a remarkable fact: the value of $\nu_s$ that gives the best fit to the two body scattering data  {\it also gives the correct triton binding energy without the introduction of any three-body forces of relativistic origin\/}.   After the initial discovery in 1996, we developed new high precision fits to the scattering data below 350 MeV lab energy, and confirmed that this remarkable conclusion persists \cite{Gross:2008ps}.  Interpretation of these results have been discussed in conference talks  \cite{Gross:1995th,trentotalk}.

The presence of off-shell couplings for scalar (and vector) mesons  will generate isoscalar interaction currents never before encountered.  Once we understood their importance, it became imperative to compute the interaction currents implied by the presence of such terms and redo the form factor calculations.  This is the purpose of this fourth generation calculation.  

\subsection{Can the CST make predictions?}

As applied to few-nucleon systems, the CST assumes that each nucleon is an off-shell composite object that can interact through two body forces constructed from OBE, with parameters fixed by fitting to $NN$ scattering data.   How can the off-shell current of the composite nucleons, and the interaction currents generated by the phenomenological $NN$ OBE interaction, be constrained?  This has been a serious issue for all models using composite hadrons and tends to undermine confidence  in such models (including the CST) to  make predictions for electromagnetic observables.  

The tool for constraining and constructing these currents is current conservation, and the general method used here was introduced by D.~O.~Riska and me   \cite{Gross:1987bu} (a similar technique, unknown by us at the time, was also developed by Ball and Chiu \cite{Ball:1980ay}).   The construction first involves finding a current for the off-shell composite nucleon that satisfies the Ward-Takahashi identity.   As reviewed in Sec.~\ref{sec:3b},  the off-shell current used here has one new off-shell nucleon form factor $F_3(Q^2)$ (subject to the constraint that $F_3(0)=1$), but its longitudinal part is otherwise completely fixed. Later in Sec.~\ref{sec:ec},  the new two-body isoscalar interaction currents implied by the off-shell, $\nu$-dependent  couplings, given in Eq.~(\ref{eq:1.2}), are constructed.    {\it These are completely new isoscalar interaction currents never before encountered\/}, and study of their effect on  the deuteron form factors, including the static magnetic and quadrupole moments, is one of the principal goals of this series of paper.  How uniquely can they be defined?

The answer to this rests on the introduction of a new concept, which I call {\it picture independence\/}.  Briefly, as emphasized in Ref.~\cite{Gross:2008ps}, {\it a
pure OBE theory with off-shell OBE couplings (picture 1) is equivalent to another theory (picture 2) with\/} no {\it off-shell OBE couplings but augmented by an infinite set of two- and three-body force diagrams constructed from meson loops that are\/} not {\it two-nucleon (or in the three nucleon sector, three-nucleon) irreducible\/}.  This equivalence might appear to be of limited use, since picture 2 involves an infinite number of irreducible kernels that are  difficult to calculate.  But, as it turns out, requiring the two pictures give equivalent electromagnetic currents places strong constraints on both.  

Of course one can always make the currents more complicated by adding purely transverse terms (i.e.~terms that satisfy the condition $q_\mu j^\mu_T=0$, where $j^\mu_T$ has no longitudinal part constrained by current conservation) but the {\it principal of simplicity\/}, as discussed in Sec.~\ref{sec:ec}, eliminates these and makes the choice of interaction current all but unique.  The interaction current used in this paper has no undetermined parameters.

This approach makes CST electromagnetic calculations fully predictable, with the possibility that they will fail.  A major goal of this fourth generation calculation, with its four related papers, is to see if the approach will work for the deuteron form factors.  One of the consequences of the principal of simplicity is that the famous $\rho\pi\gamma$ exchange current is excluded from consideration.  Adding such a current (and its $\omega\sigma\gamma$ companion) \cite{Hummel:1989qn} can certainly have some effect, particularly if unrealistic assumptions are made about their structure \cite{Chemtob:1974nf}; these currents certainly contribute at some level of accuracy.   In Ref.~II I will show that the  interaction currents fixed in this paper give a result for the magnetic moment that is very close to the experimental data, indicating that all other contributions are very small indeed.

\subsection{Summary of the paper}

The next section (Sec.~\ref{sec:II}) defines the $dnp$ vertex function and the relativistic kernel used in the CST bound state equation, and reviews the work of Ref.~\cite{Gross:1987bu}, which showed that the exact result for the form factors in the CST requires the calculation of only four diagrams (shown in Fig.~\ref{Fig1}), one of which is the interaction current contribution.  If the interaction current  satisfies the two-body Ward-Takahashi (WT) identity derived in Appendix \ref{app:WTidet}, the two body current is conserved, provided the one body current satisfies the one-body WT identity.  The section concludes by discussing the covariant normalization condition, and showing that the normalization condition guarantees that the deuteron charge is unity, results previously reported in Refs.~\cite{Gross:1969rv,Gross:1987bu,Adam:1997rb,Adam:1997cx}.  

As mentioned above,  Sec.~\ref{sec:ec} uses the principals of simplicity and picture independence to derive the isoscalar interaction current.  This derivation is one of the principal new results of this paper.  The interaction current will be shown to factor into a sum  of products of the nucleon current multiplied by truncated $NN$ interaction kernels  constructed from the full OBE $NN$  kernel.  Only scalar and vector meson exchanges contribute; contributions from $\Theta$ terms  present in the pseudoscalar exchanges cancel.  

The factorized form of the interaction current permits these contributions to be expressed as a product of the nucleon current and truncated deuteron wave functions, which can be combined with the CIA contributions.  The interaction current contributions from the off-shell particle 2 are represented by a new, truncated wave function, $\Psi^{(2)}$, and combine with the contributions from diagram (A) of Fig.~\ref{Fig1}.  Contributions from the (originally on-shell) particle 1 cancel some of the contributions from the (B) diagrams; their combined contributions can be expressed in terms of a new wave function, $\widehat \Psi$.   Section~\ref{sec:ec} concludes with general formulae for the deuteron current given as traces over products of the wave (or vertex) functions, propagators, and nucleon currents, the second principal result of this paper.

Evaluation of the results for the magnetic moment is the main result of the second paper in this series, Ref.~II, prepared at the same time as this paper.  Calculation of the magnetic moment is the first numerical prediction of the consequences of the interaction currents found in this work.

\section{Deuteron current} \label{sec:II}

The structure of the composite deuteron enters through the $dnp$ vertex function and covariant wave function.  In this section the notation for the vertex function, the wave function, and the CST bound state equations that they satisfy is given.  The bound state equation was solved in Ref.~\cite{arXiv:1007.0778}, and the wave functions used in Ref.~II were obtained there.  Following this, the general formulae for the deuteron current is given, the role of the strong nucleon form factors discussed, and the diagrams that do not depend on the interaction current  reduced.  The section concludes with derivation of the two-body WT identity that the interaction current must satisfy and a review of the normalization condition for the deuteron wave functions and its connection to the deuteron charge.

\subsection{Deuteron vertex and wave functions}

The relativistic structure of the deuteron bound state is written in terms of the covariant $dnp$ vertex function.  For an incoming deuteron of four momentum $P$ and polarization four-vector $\xi$, this vertex function is written
\bea
{\cal G}_{\alpha\beta}^{\lambda_d}(k,P)&=&(\Gamma_\nu{\cal C})_{\alpha\beta}(k,P)\xi^\nu_{\lambda_d}(P)
\nonumber\\
&=&\Gamma_{\alpha\beta'}^{\lambda_d}(k,P){\cal C}_{\beta'\beta}\, ,
\label{eq:2.1a}
\eea
where $k$ is the four-momentum of particle 1 (with Dirac index $\beta$) and   $p=P-k$ is the four-momentum of particle 2 (with Dirac index $\alpha$) and $\lambda_d$ is the polarization of the deuteron.  Since $p$ is not an independent variable, it will usually be suppressed, except when we need to refer to the momentum of particle 2 explicitly, in which case we will sometimes use $p$ instead of $P-k$.   This vertex function is sometimes needed when {\it both} particles are off-shell (when it will sometimes carry the subscript ``BS'' to avoid confusion), but when one particle is on-shell it will always be particle 1, and in this case $k^2=m^2$, and we may sometimes use the on-shell matrix element
\bea
{\cal G}_{\alpha\lambda}^{\lambda_d}(k,P)&=&{\cal G}_{\alpha\beta}^{\lambda_d}(k,P)\,\bar{u}^T_\beta({\bf k},\lambda)\, ,
\label{eq:2.1}
\eea
distinguished from ${\cal G}_{\alpha\beta}$ {\it only\/} by the replacement of the Dirac index $\beta$ by the nucleon helicity index $\lambda$.
The matrix element (\ref{eq:2.1}) has the structure of a nucleon spinor; it describes the state of an incoming deuteron and two outgoing nucleons, one of which is off-shell.  Note that the convention for writing the vertex function ${\cal G}$ in this paper differs from Eq.~(3.7) of Ref.~\cite{arXiv:1007.0778}.  Here the indices $\alpha$ and $\beta$ {\it are interchanged\/} and the spinor of the on-shell particle 1 is contracted from the right.  This is done to simplify the formulae for the current, but all results are independent of this change.

The state of an outgoing deuteron is obtained by taking the Dirac conjugate, defined by
\bea
\overline{{\cal O}}=\gamma^0{\cal O}^\dagger \gamma^0\, .
\eea
For the general case, the conjugate vertex function is therefore
\bea
\overline{\cal G}^{\lambda_d}(k,P)&=&\gamma^0{\cal G}^{\lambda_d\,\dagger}(k,P)\gamma^0
\nonumber\\
&=&{\cal C}\gamma^0\Gamma^\dagger_\nu(k,P)\gamma^0\,\xi^{\nu\,*}_{\lambda_d}(P)
\nonumber\\
&=&{\cal C}\,\overline{\Gamma}_\nu(k,P)\xi^{\nu\,*}_{\lambda_d}(P)
\, ,
\eea
where use was made of $\gamma^0 {\cal C}^\dagger \gamma^0={\cal C}$.  
Using the properties of  $\gamma^0$,
\bea
\gamma^0 (\gamma^\mu)^\dagger \gamma^0&=&\gamma^\mu
\eea
the outgoing vertex function can be written in a more intuitive form.  Consider a term in $\Gamma_\nu$ of the form $\gamma_\nu \slashed{P}$ and go through the steps explicitly:
\bea
\gamma^0\Gamma_\nu^\dagger\gamma^0&\to&\gamma^0\slashed{P}^{\dagger}\gamma_\nu^\dagger\gamma^0=\slashed{P}\gamma_\nu
\to \overline{\Gamma}_\nu\, .\qquad
\label{eq:2.5}
\eea
We therefore recognize that $\overline{\Gamma}_\nu$ is the matrix $\Gamma_\nu$ with its $\gamma$ matrices {\it written in reverse order\/} (and the deuteron polarization vector complex conjugated).

If particle 1 is on-shell, (\ref{eq:2.1}) and (\ref{eq:2.5}) imply that
\bea
\overline{\cal G}_{\lambda\alpha}^{\lambda_d}(k,P)&=&{u}^T_\beta({\bf k},\lambda)({\cal C}\,\overline{\Gamma}_\nu)_{\beta\alpha}(k,P) \xi^{\nu\,*}_{\lambda_d}(P)
\nonumber\\
&=&{u}^T_\beta({\bf k},\lambda)\,\overline{\cal G}_{\beta\alpha}^{\lambda_d}(k,P) . \label{eq:spinleft}
\eea
The on-shell particle is now on the left, the correct location for constructing matrix elements.

It is convenient to introduce covariant wave functions.  Two different wave functions will be defined. When the charge conjugation matrix is included, the notation is 
\bea
{\it \Psi}_{\alpha\beta}^{\lambda_d}(k,P)&=&S_{\alpha\alpha'}(p){\cal G}_{\alpha'\beta}^{\lambda_d}(k,P)
\nonumber\\
\overline{\it \Psi}_{\beta\alpha}^{\lambda_d}(k,P)&=&\overline{\cal G}_{\beta\alpha'}^{\lambda_d}(k,P){S}_{\alpha'\alpha}(p)\, ,\label{eq:27}
\eea
where $S$ is the {\it bare\/} nucleon propagator (see the discussion in Sec.~\ref{sec:3b} below).   Sometimes it is convenient to remove the charge conjugation matrix explicitly from both sides of the relations (\ref{eq:27}), giving
\bea
{\Psi}_{\alpha\beta}^{\lambda_d}(k,P)&=&S_{\alpha\alpha'}(p){\Gamma}_{\alpha'\beta}^{\lambda_d}(k,P)
\nonumber\\
\overline{\Psi}_{\beta\alpha}^{\lambda_d}(k,P)&=&\overline{\Gamma}_{\beta\alpha'}^{\lambda_d}(k,P){S}_{\alpha'\alpha}(p)\, .\label{eq:27a}
\eea
The reader is cautioned to be aware of the difference between ${\it \Psi}$ and $\Psi$, related the the charge conjugation factor ${\cal C}$
\bea
{\it \Psi}^{\lambda_d}(k,P)= \Psi^{\lambda_d}(k,P)\,{\cal C} \, . \label{eq:twopsis}
\eea
In both cases, the replacement of the index $\beta$ by $\lambda$ denotes multiplication from the right [as in Eq.~(\ref{eq:2.1})] or the left by the transpose  [as in Eq.~(\ref{eq:spinleft})]  of the appropriate on-shell nucleon spinor.

Most of the results of this  paper are expressed in terms of the manifestly covariant  $dnp$ vertex and wave functions, Eqs.~(\ref{eq:2.1a}), (\ref{eq:27}), and their various alternate forms.  These functions can be expressed in terms invariant functions, and also in terms of  the familiar nonrelativistic S and D-state wave functions, $u$ and $w$ and the small P-state components, $v_t$ and $v_s$.  Use of these expansions is postponed until Ref.\ II.

\subsection{Equation for the bound state wave function}

The equation for the bound state wave function (\ref{eq:27}) with particle 1 on shell  (c.f. Eq.~(3.7) of Ref.~\cite{arXiv:1007.0778}, with the notational change in ${\it \Psi}$ mentioned above) is
\bea
S^{-1}_{\alpha\alpha'}(p)&&{\it \Psi}^{\lambda_d}_{\alpha'\lambda}(k,P)
\nonumber\\
&&=-\int_{k'} \overline V_{\lambda\lambda',\alpha\alpha'}(k,k';P){\it \Psi}_{\alpha'\lambda'}(k',P)\, , \qquad \label{eq:bsequation}
\eea
where $\overline{V}$ is the symmetrised kernel (introduced in Ref.~\cite{Gross:2008ps} and discussed below) and the volume integral is
\bea
\int_{k}=\int \frac{d^3 k}{(2\pi)^3}\frac{m}{E_k}\, .
\label{eq:volint1}
\eea
Here particle 1, with four momentum $k=\{E_k,{\bf k}\}$, is on shell (so that $E_k=\sqrt{m^2+{\bf k}^2}$).

It is sometimes convenient to remove the on-shell spinors and write this equation in terms of Dirac matrices only.  The sum over the polarization of the on-shell spinors can be replaced using the  positive energy Dirac projection operator  
\bea
\Lambda_{\gamma\gamma'}(k)=\frac{(m+\slashed{k})_{\gamma\gamma'}}{2m}=\sum_{\lambda'} {u}_{\gamma}({\bf k},\lambda')\bar{u}_{\gamma'}({\bf k},\lambda') ,\qquad \label{eq:spindecom}
\eea
and, recalling the transpose that appears in Eq.\ (\ref{eq:2.1}), this gives
\bea
&&S^{-1}_{\alpha\alpha'}(p){\it \Psi}^{\lambda_d}_{\alpha'\beta}(k,P)
\nonumber\\
&&\qquad=-\int_{k'} \overline V_{\beta\gamma,\alpha\alpha'}(k,k';P){\it \Psi}_{\alpha'\gamma'}(k',P) \Lambda^T_{\gamma'\gamma}(k)\, , \qquad\quad
\eea
This equation can be decomposed into partial waves as discussed in Ref.~\cite{arXiv:1007.0778} and reviewed in Ref.\ II.  For this paper  the partial wave amplitudes are not needed.

\begin{figure*}
\centerline{
\mbox{
\includegraphics[width=5.6in]{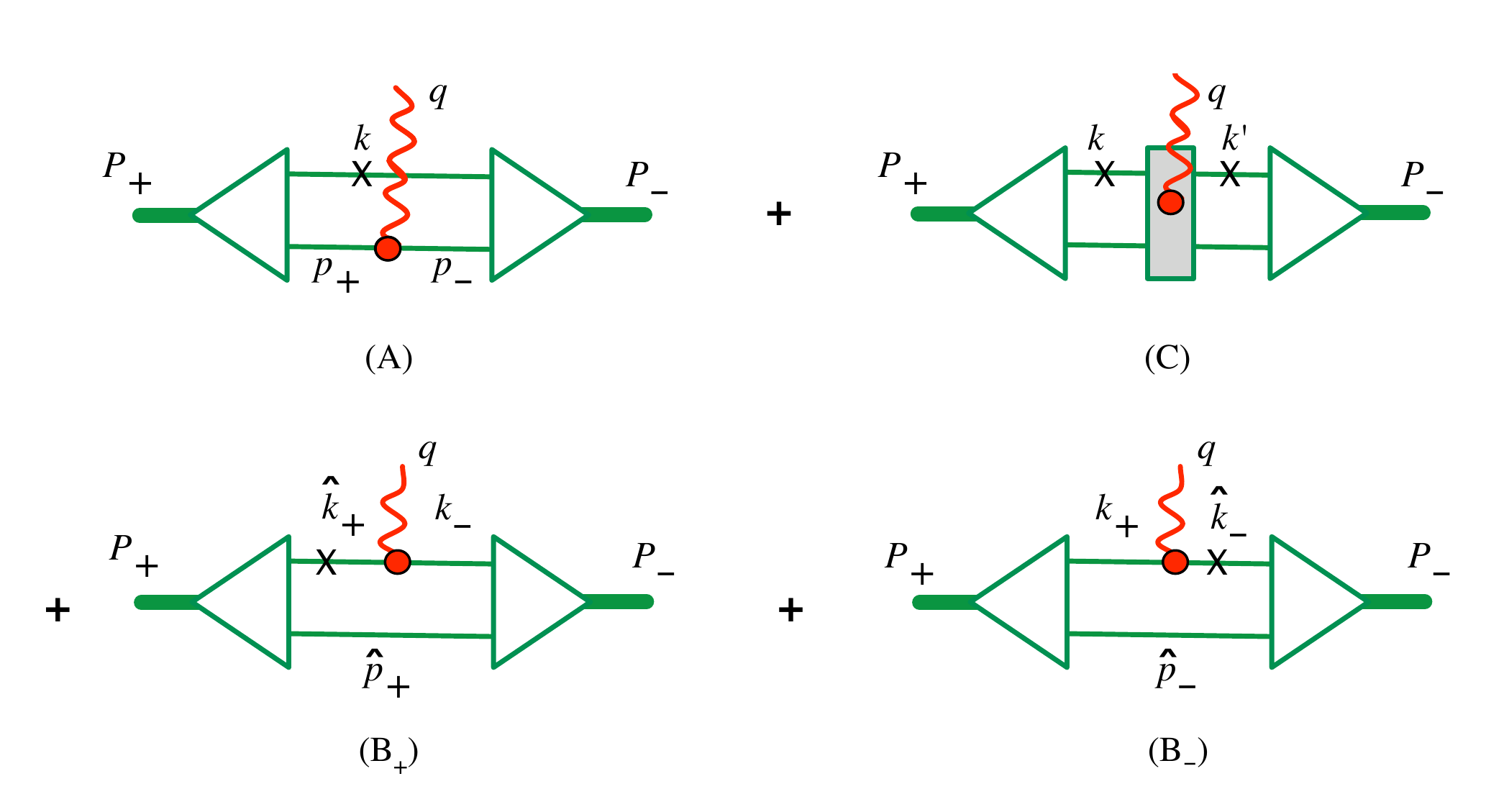}
}
}
\caption{\footnotesize\baselineskip=10pt (Color on line) Diagramatic representation of the current operator of the Covariant Spectator Theory with particle 1 on-shell (the on-shell particle is labeled with a $\times$).  Diagrams (A), (B$_+$), and (B$_-$) are the complete impulse approximation (CIA), while (C) is the interaction current term.  In some cases the sum of the (B) diagrams is approximately equal to the first, and for this reason the relativistic impulse approximation (RIA) is the defined to be two times the first term, (A).
}
\label{Fig1}
\end{figure*} 

\subsection{General formula for the bound state current}

As discussed in Ref.~\cite{Gross:1987bu}, the CST two-body current operator can be expressed in terms of the four diagrams shown in Fig.~\ref{Fig1}.  Using the notation for the vertex and wave functions reviewed above, they can be written
\begin{widetext}
\begin{align}
J^\mu_{\lambda\lambda'}(q)=\int_k \Bigg\{&
 \overline{\it \Psi}_{\lambda_n\alpha}^{\lambda}(k,P_+)\,j^\mu_{\alpha\alpha'}(p_+,p_-)\,{\it \Psi}_{\alpha'\lambda_n}^{\lambda'}(k,P_-)+
 \int_{k'}
 \overline{\it \Psi}_{\lambda_n\alpha}^{\lambda}(k,P_+)\,V^\mu_{\lambda_n\lambda_n',\alpha\alpha'}(k\,P_+; k'\,P_-)\,{\it \Psi}_{\alpha'\lambda_n'}^{\lambda'}(k',P_-)  
\nonumber\\
 +&\left[\frac{E_k}{E_+}\right] \overline{\it\Psi}_{\lambda_n\alpha}^{\lambda}(\hat k_+,P_+)\,{\cal G}_{\alpha\beta}^{\lambda'}(k_-,P_-)\Big[\bar u_\gamma({\bf k}_+,\lambda_n) \,j^\mu_{\gamma\gamma'}(\hat k_+, k_-)\,S_{\gamma'\beta}(k_-)\Big]^T
\nonumber\\&
+\left[\frac{E_k}{E_-}\right] \overline{\cal G}_{\beta\alpha'}^{\lambda}(k_+,P_+)\,{\it\Psi}_{\alpha'\lambda_n'}^{\lambda'}(\hat k_-,P_-)\Big[S_{\beta\gamma}(k_+) \,j^\mu_{\gamma\gamma'}(k_+, \hat k_-)\,u_{\gamma'}({\bf k}_-,\lambda_n') \Big]^T\Bigg\}\label{eq:23}
\end{align}
\end{widetext}
where sums over repeated polarization indicies are implied,  $V^\mu$ is the two-nucleon interaction current,  and
\bea
P_\pm&&=D\pm \sfrac12 q
\eea
with (in the Breit frame) $D=\{D_0, {\bf 0}\}$, and $q=\{0, {\bf q}\}$ with ${\bf q}=\{0,0,q_z\}$ and $D_0=\sqrt{m_d^2+\frac14 {\bf q}^2}$.  In diagrams A and C, $k$ and $k'$ are on shell, so that $k^2=k'^2=m^2$ and $k=\{E_k,{\bf k}\}$ and $k'=\{E_{k'},{\bf k}'\}$.  
In diagram A,
\bea
p_\pm&&=P_\pm- k\, . \label{eq:pppm}
\eea
In the last two diagrams, B$_\pm$, the hat is used to label the on-shell four-momentum, so that 
\bea
\hat k_\pm &=& \{E_\pm, {\bf k}\pm\sfrac12{\bf q}\}
\nonumber\\
k_\pm &=& \{E_\mp, {\bf k}\pm\sfrac12{\bf q}\}
\nonumber\\
\hat p_\pm&=&P_\pm-\hat k_\pm=P_\mp-k_\mp\, , \label{eq:khatkp}
\eea
with $E_\pm=\sqrt{m^2+({\bf k}\pm\frac12{\bf q})^2}$.  All four terms have been written using the same volume integral (\ref{eq:volint1}),
but, in the last two terms, this integral must be corrected since in the B$_\pm$ terms the correct on-shell energy is $E_\pm$ and not the $E_k$ included in the volume integral.  Note that the vertex functions ${\cal G}$ and $\overline{\cal G}$ in the last two terms have both particles off-shell.

The single nucleon current is an operator in isospin space
\bea
j^\mu(q)= j^\mu_s(q)+ j^\mu_v(q)\,\tau_3\to j_s^\mu(q)\to j^\mu(q)\, ,
\eea
since in this paper we consider the isoscalar currents only (and drop the subscript $s$).  This current is sufficient for study of the deuteron form factor, the focus of this paper.  At $q=0$, the current reduces to
\bea
j^\mu(0)=\frac12=e_0\, . \label{eq:e0}
\eea
where $e_0$ is the isoscalar charge (consistent with Eq.~(2.17) of Ref.~\cite{Gross:1987bu}).  Even though $e_0=\frac12$, we will continue to use $e_0$ throughout this paper.

\subsection{Role of the strong nucleon form factor}\label{sec:3b}

In the CST calculations under discussion, the $NN$ scattering kernel includes a {\it strong\/} nucleon form factor, which we will denote by $h(p)$ in this paper \cite{Adam:2002cn} (this form factor is a scalar function of $p^2$ but written simply as $h(p)$ for convenience).  This strong form factor is related to the nucleon self energy, and is to be distinguished from the {\it electromagnetic\/} form factors of the nucleon (the connection between them is discussed below).  The strong form factor $h(k)=1$ when the nucleon is on-shell ($k^2=m^2$).    To avoid confusion with the electromagnetic form factors of the nucleon, we will always refer to $h$ as the {\it strong\/} form factor.  

In the OBE models that have been used, where the strong form factors at the meson-$NN$ vertices are  products of strong from factors for each particle entering or leaving the vertex, the strong form factor  associated with each external nucleon line can be removed from the $NN$ scattering kernel and the interaction current, leading to
\bea
&&\overline{V}(k,k';P)=h(k)h(p)\widetilde V(k,k';P)h(k')h(p')
\nonumber\\
&&V^\mu(kP_+;k'P_-)= h(p_+)\widetilde V^\mu(kP_+;k'P_-)h(p'_-) \qquad \label{eq:VandVmu}
\eea
where $\widetilde V$ is the {\it reduced\/} kernel and  $\widetilde V^\mu$ the {\it reduced\/} interaction current, and we recall that, for both primed and unprimed variables, $p=P-k$ and $p_\pm=P_\pm-k$.  Note that the expression for the kernel is written allowing for the possibility that any (or all four) of the particles could be off-shell, but the expression for the interaction current, $V^\mu$, assumes that $k^2=k'^2=m^2$.  

When proving current conservation it is best to remove the strong form factors from the interaction kernel and interaction current, and incorporate them into the propagator, giving a dressed propagator of the form
\bea
S_d^{-1}(p)&=&\frac{m-\slashed{p}}{h^2(p)}=\frac{S^{-1}(p)}{h^2(p)} \, ,\label{eq:Sd}
\eea
where the $h$ occurs squared because one comes from the initial and one from the final interactions that connect the propagator.  
Following the method of Riska and Gross \cite{Gross:1987bu}, a conserved two-nucleon current can then be constructed using a {\it dressed\/} single nucleon current of the form \cite{Adam:2002cn}
\bea
j^\mu(p,p')=h(p)h(p')j_R^\mu(p,p')\, . \label{eq:3.4}
\eea
where  the {\it reduced\/} current $j_R^\mu$  satisfies the Ward-Takahashi (WT) identity
\bea
q_\mu \, j_R^\mu(p,p')&&=e_0\Big[S^{-1}_{d}(p')-S^{-1}_{d}(p)\Big]
\, , \label{eq:210a}
\eea
where the isoscalar charge, $e_0$, was introduced in  Eq.~(\ref{eq:e0}).  The WT identity for the dressed current is then
\bea
q_\mu \, j^\mu(p,p')&&=e_0\Big[\frac{h(p)}{h(p')}S^{-1}(p')-\frac{h(p')}{h(p)}S^{-1}(p)\Big]\, .\qquad
\, , \label{eq:WTdressed}
\eea

The development depends very strongly on whether of not there is a strong form factor $h$ different from unity.  To see the connection between the current and $h$, it is sufficient to look at the WT identity for the nucleon current, Eq.\ (\ref{eq:210a}), at small $q$ and expand the right-hand side.  This gives the condition
\bea
j_R^\mu(p,p)=-e_0 \frac{\partial S^{-1}_d(p)}{\partial p_\mu}.
\eea
If $h=1$, so that $S_d^{-1}(p)\to S^{-1}(p)=m-\slashed{p}$, this gives the familiar current
\bea
j_R^\mu(p,p)\to j^\mu(p,p)=e_0\,\gamma^\mu \, .
\eea
However, using the dressed propagator (\ref{eq:Sd}) gives the result
\bea
j_R^\mu(p,p)=\frac{e_0}{h^2}\Big\{\gamma^\mu+\frac{4m}{h}\frac{\partial h}{\partial p_\mu}\Theta(p)\Big\}\, ,\label{eq:33}
\eea
where $\Theta(p)$ was defined in Eq.~(\ref{eq:theta}) 
and, for the full current
\bea
j^\mu(p,p)=e_0\Big\{\gamma^\mu+\frac{4m}{h}\frac{\partial h}{\partial p_\mu}\Theta(p)\Big\}\, .\label{eq:33a}
\eea

Thus, when $h\ne1$, two  completely equivalent descriptions are possible.  One may use {\it either\/}

\begin{enumerate}
\item the reduced current $j^\mu_R$, the dressed propagator $S_d$, and the reduced interactions $\widetilde V$ and $\widetilde V^\mu$, or

\item the full current $j^\mu$, the bare propagator $S$, and the full interactions $\overline{V}$ and $V^\mu$.  

 \end{enumerate}

In the following discussion we will sometimes remove the strong form factors from the {\it kernel and the interaction current\/} (relying in the difference between $\overline{V}$ and $\widetilde{V}$ to distinguish between the two), but will use the {\it bare\/} propagator $S$ and always include the strong form factors in the vertex functions and the wave functions, where they occur naturally [consistent with the definitions (\ref{eq:27}) in which the propagator $S$ that appears is the bare propagator].  We will use the full current $j^\mu$.  These conventions will require that, in cases when $\widetilde{V}$ is used in place of $\overline{V}$, the $h$ are {\it written explicitly\/}.  This notation has the advantage that Eq.~(\ref{eq:23}) for the two-nucleon current is unchanged in the presence of the strong nucleon form factor.

\subsection{General form of the nucleon current with $h\ne1$}\label{subs:E}

To prepare for a general discussion of the charge and normalization, we summarize previous results for the general form of the off-shell nucleon current when $h\ne1$.

The simplest from of off-shell nucleon current consistent with current conservation (in the notation of 
Ref.~\cite{Gilman:2001yh}) is
\bal
 j^\mu(p',p)&=e_0\,f_0(p',p)\gamma^\mu+e_0\,g_0(p',p)\Theta(p')\gamma^\mu\Theta(p)
\nonumber\\
&\quad+e_0\,f_0(p',p)\left\{(F_1-1)\widetilde\gamma^\mu+F_2\frac{i\sigma^{\mu\nu}q_\nu}{2m}\right\} \qquad
\nonumber\\
&\quad+ e_0\,g_0(p',p)\Theta(p')(F_3-1)\widetilde\gamma^\mu\Theta(p) \label{3.1}
\end{align}
where $q=p'-p$, $F_i=F_i(q^2)$ are the form factors of the nucleon (with $F_3$, subject to the constraint that $F_3(0)=1$, a new form factor that contributes only when both nucleons are off-shell), and the transverse gamma matrix is
\begin{equation}
\widetilde\gamma^\mu=\gamma^\mu-\frac{\slashed{q}q^\mu}{q^2}\, . \label{eq:gammatilde}
\end{equation}

Note that the terms involving the form factors $F_i$ are all transverse.  The functions $f_0$ and $g_0$ describe the modification of the current due to the presence of the strong form factor $h$.  Using the shorthand notation $h=h(p)$ and $h'=h(p')$, these functions are
\bea
f_0(p',p)&=&\frac{h'}{h} \frac{(m^2-p^2)}{p'^2-p^2}+\frac{h}{h'} \frac{(m^2-p'^2)}{p^2-p'^2}
\nonumber\\
g_0(p',p)&=&\frac{4m^2}{p'^2-p^2}\left(\frac{h}{h'}-\frac{h'}{h}\right)
\, .
\eea
These functions are consistent with the Ward--Takahashi identity (\ref{eq:210a}), and both are regular at $p'^2=p^2$.  Note the  useful relation
\bea
f_0(p',p)=&&\frac12\Big[\frac{h'}{h}+\frac{h}{h'}\Big]
\nonumber\\&&-\frac{g_0(p',p)}{8m^2}(2m^2-p'^2-p^2).\qquad\quad \label{eq:fgrelation}
\eea

Several limits of this current are interesting.  First, when either the initial or final particle is on shell, the term involving $F_3$ will not contribute, and the current reduces to
\bea
\lim_{p^2=m^2}j^\mu(p',p)&\to& \frac{1}{h'}  j_0^\mu (q)
\nonumber\\
\lim_{p'^2=m^2}j^\mu(p',p)&\to& \frac{1}{h}  j_0^\mu (q) \label{eq:F0simplify}
\eea
where 
\bea
j^\mu_0(q)=e_0\left\{\gamma^\mu + (F_1-1)\widetilde\gamma^\mu+F_2\frac{i\sigma^{\mu\nu}q_\nu}{2m}\right\}\, . \label{eq:j0current}\qquad
\eea
Note that the simplified current $j^\mu_0$ satisfies a simplified WT identity
\bea
q_\mu j_0^\mu (q)=e_0[S^{-1}(p')-S^{-1}(p)]=e_0\,\slashed{q}\, . \label{eq:WTforj0}
\eea

Secondly, when $q\to0$ (but neither particle is on shell)  the current reduces to
\bea
j^\mu(p,p)=e_0\,f_{00}\,\gamma^\mu+e_0\,g_{00}\,\Theta(p)\gamma^\mu\Theta(p)\, ,\label{eq:jat0}
\eea
where 
\bea
f_{00}\equiv\lim_{{p'^2\to p^2}} f_0(p',p)&=&1+2a(p^2)(m^2-p^2)\nonumber\\
g_{00}\equiv\lim_{{p'^2\to p^2}} g_0(p',p)&=&-8m^2\,a(p^2) \label{eq:FGexpand}
\eea
with 
\bea
a(p^2)=\frac{1}{h}\frac{dh}{dp^2}\, . \label{4.7}
\eea
Finally, in Sec.~\ref{sec:combined} it will be convenient to observe the the full current can be written in two convenient forms
\bea
j^\mu(p',p)&=&f_0(p',p)j_0^\mu(q)+\Theta(p')\,j_{\rm off}^\mu(p',p)\,\Theta(p)
\nonumber\\
&=&\frac12\Big[\frac{h'}{h}+\frac{h}{h'}\Big]j^\mu_0(q)+j^\mu_g(p',p)
\qquad \label{eq:fulljsimp}
\eea
where $j_{\rm off}^\mu$ contributes {\it only\/} when both particles are off-shell, and $j^\mu_g$ is the part of the current that depends on $g_0$
\bea
j_{\rm off}^\mu(p',p)&\equiv& e_0\,g_0(p',p)\Big[(F_3-1)\tilde\gamma^\mu+\gamma^\mu\Big]
\nonumber\\
j^\mu_g(p',p)&=&-\frac{g_0(p',p)}{8m^2}(2m^2-p'^2-p^2)\,j^\mu_0(q)
\nonumber\\
&&+\Theta(p')\,j_{\rm off}^\mu(p',p)\,\Theta(p)\, .\qquad
\eea

The full form of the current 
(\ref{eq:fulljsimp}) 
is needed only for the proof of current conservation.  In all applications (after current conservation has been proved) the terms proportional to $q^\mu$ can be dropped because either the helicity amplitudes are needed (and $q\cdot\epsilon=0$), or the nucleon current will be coupled to another conserved current.  
Dropping the $q^\mu$ terms reduces the full current (\ref{eq:fulljsimp}) and the simplified current (\ref{eq:j0current})  to
\bal
 j^\mu(p',p)\to&\,f_0(p',p) j^\mu_{N}(q)
\nonumber\\
&+e_0\, g_0(p',p)\Theta(p')F_3(q^2)\gamma^\mu\Theta(p)
\nonumber\\
 j^\mu_0(q)\to&\, j^\mu_{N}(q)
  \label{3.8b}
\end{align}
where $j^\mu_{N}(q)$ is the familiar isoscalar nucleon current (with $e_0$ included)
\bea
 j^\mu_{N}(q)=e_0\,F_1(q^2)\gamma^\mu+e_0\,F_2(q^2)\frac{i\sigma^{\mu\nu}q_\nu}{2m}\, . \label{eq:jN}
\eea
The conditions (\ref{eq:FGexpand}) insure that $j^\mu(p,p')\to j_N^\mu(q)$ when both nucleons are on shell.

The interesting off-shell term proportional to the new nucleon form factor, $F_3(q^2)$, will contribute only to diagram \ref{Fig1}(A), because the (B) diagrams have either the initial or final nucleon on shell, and  one of the $\Theta$ projection operators will always vanish.   Under these conditions the nucleon current reduces to (\ref{eq:F0simplify}) insuring that, in the (B) diagrams where particle 1 may have off-shell momenta $k_\pm$, the factors of $h(k_\pm)$ contained in the vertex functions will cancel.

\subsection{Reduction of the (A) and (B) diagrams}\label{subs:F}

In doing calculations, it is convenient to write the (A) and (B) diagrams as traces over products of $\gamma$ matrices.  

To rewrite the (A) diagram  as a trace, recall that the incoming and outgoing vertex functions both  have a factor of ${\cal C}$ [recall Eqs.~(\ref{eq:2.1a}) and (\ref{eq:spinleft})] which is also contained in the wave function ${\it \Psi}$ [recall (\ref{eq:twopsis}
)].  Use (\ref{eq:spindecom}) to replace the sum over the positive energy spinors and extract and remove the charge conjugation matrices, recalling that ${\cal C}=-{\cal C}^{-1}$ and using  
\bea
{\cal C}^{-1}\gamma^{\mu\,T}{\cal C}&=&-\gamma^\mu\, .
\label{eq:crelations1}
\eea
The result is
\bea
&&J^\mu_{\lambda\lambda'}(q)\Big|_{{\rm A}}
\nonumber\\
&&=-\int_k {\rm tr}\Big[\overline{\Psi}^\lambda (k,P_+)j^\mu(p_+,p_-)
\Psi^{\lambda'}(k,P_-)\,\Lambda(-k)\Big], \quad\qquad\label{eq:Atrace}
\eea
where the wave functions $\Psi$ will be written an terms of $\gamma$ matrices and invariants Ref.\ II.

The B$_\pm$ terms in Eq.~(\ref{eq:23}) are each singular as $q_\mu\to0$, but together are finite.  They can be combined conveniently if we note that the two off-shell propagators (where the singularity is located) are
\bea
S_{\alpha \beta}(k_\pm)&=&\frac{2m\Lambda_{\alpha\beta}(k_\pm)}{m^2-k_\pm^2}
\nonumber\\&=&
\frac{2m\Lambda_{\alpha\beta}(k_\pm)}{E_\pm^2-E_\mp^2}=\frac{2m\Lambda_{\alpha\beta}(k_\pm)}{\pm2{\bf k}\cdot{\bf q}}\, , \label{eq:Skpkm}
\eea
where $S(k_+)$ is the propagator at the pole where $k_0=E_-$.  
Hence, restoring the positive energy projection operator using (\ref{eq:spindecom}),  the two B$_\pm$ terms combine into a symmetric form.  Introducing the {\it reduced\/} vertex functions (for the general case when both $k$ and $p$ are off-shell) 
\bea
{\cal G}(k,P)={\Gamma}(k,P)\,{\cal C}=h(k)h(p)\,\widetilde{\Gamma}(k,P)\, {\cal C} \label{eq:reducedGs}
\eea
and extracting the charge conjugation operator as was done for the (A) diagrams gives
\begin{widetext}
%
\begin{align}
J^\mu_{\lambda\lambda'}(q)\Big|_{{\rm B}_\pm}
=\int_k\left[\frac{m E_k}{{\bf k}\cdot{\bf q}}\right]{\rm tr} \Bigg\{&\frac1{k_0}\,
\overline{\widetilde\Gamma}^{\lambda}(\widetilde k_+,P_+)\, \Big[h^2(\widetilde p)S(\widetilde p)\Big]\,\widetilde\Gamma^{\lambda'}(\widetilde k_-,P_-)\Lambda(-\widetilde k_-) \,j_0^\mu(q)\,\Lambda(-\widetilde k_+) \Big|_{k_0=E_-}
\nonumber\\& 
-\frac1{k_0}\, \overline{\widetilde\Gamma}^\lambda(\widetilde k_+,P_+)\, \Big[h^2(\widetilde p)S(\widetilde p)\Big]\,\widetilde\Gamma^{\lambda'}(\widetilde k_-,P_-) \Lambda(-\widetilde k_-) \,j^\mu_0(q)\,\Lambda(-\widetilde k_+)\Big|_{k_0=E_+} \Bigg\}\, ,
\label{eq:Bpmterms3}
\end{align}
\end{widetext}
where  the notation   
\bea
\widetilde k=\{k_0,{\bf k}\} ,\qquad
\widetilde k_\pm= \widetilde k \pm \sfrac12 q, \qquad
\widetilde p=D-\widetilde k\, ,\qquad \label{eq:ktilde}
\eea
[recall that $D=\frac12(P_++P_-$)] is used to display the fact that the two terms are identical except for the energy factors $E_\pm$.  As $q\to0$, $E_-\to E_+$ showing that the numerator approaches zero and the singularity is cancelled.    To reduce the two (B) terms to 
(\ref{eq:Bpmterms3}) it is necessary to use (\ref{eq:F0simplify}),  the behavior of the current under charge conjugation 
\bea
{\cal C}^{-1}j^{\mu\,T}(k',k){\cal C}&=&-j^\mu(-k,-k')
\nonumber\\
{\cal C}^{-1}j_0^{\mu\,T}(q){\cal C}&=&-j_0^\mu(q)\, , \label{eq:crelations}
\eea
and to recall that the sign introduced through the substitution ${\cal C}=-{\cal C}^{-1}$ is cancelled by the change is the sign of the current.   Note that the off-shell strong from factors for particle 1 that were originally contained in ${\cal G}(k_-,P_-)$ and ${\cal G}(k_+,P_+)$ are cancelled by the strong from factor that arises in the conversion of $j^\mu\to j^\mu_0$ [recall Eq.~(\ref{eq:F0simplify})], so that all strong form factors for particle 1 are removed from these diagrams, even thought particle 1 is {\it off-shell in parts of these diagrams\/}.  These diagrams also do  not include contributions from $F_3$.

\subsection{Two-body Ward-Takahashi identity}

The condition that the reduced interaction current $\widetilde{V}^\mu(kP_+;k'P_-)$ must satisfy in order that the total current, $J^\mu(q)$ be conserved is referred to as the two-body WT identity, and is derived in Appendix \ref{app:WTidet}.  It can be written
\begin{align}
q_\mu & \widetilde{V}^\mu_{\beta\beta',\alpha\alpha'}(k\,P_+;k'\,P_-)
\nonumber\\
=&e_0\Big[ \widetilde{V}_{\beta\beta',\alpha\alpha'}(k,k';P_-)- \widetilde{V}_{\beta\beta',\alpha\alpha'}(k,k';P_+)
 \nonumber\\
 &+  \widetilde{V}_{\beta\beta',\alpha\alpha'}(k-q,k';P_-)- 
 \widetilde{V}_{\beta\beta',\alpha\alpha'}(k_,k'+q;P_+)\Big] . \label{eq:29}
\end{align}
When the interaction current is constructed below, it will be shown to satisfy (\ref{eq:29}) thus insuring that the total current is conserved.

Expanding the right hand side of (\ref{eq:29}) in powers of $q$ and retaining the linear term only, gives
\begin{align}
\lim_{q\to0}& \widetilde{V}^\mu_{\beta\beta',\alpha\alpha'}(k\,P_+;k'\,P_-)
=-2e_0\frac{\partial}{\partial P_\mu}\widetilde{V}_{\beta\beta',\alpha\alpha'}(k,k';P)
 \nonumber\\
&-e_0\frac{\partial}{\partial k_\mu} \widetilde{V}_{\beta\beta',\alpha\alpha'}(k,k';P)
-e_0 \frac{\partial}{\partial k'_\mu} 
\widetilde{V}_{\beta\beta',\alpha\alpha'}(k,k';P)  \label{eq:211}
\end{align}
%
where the partials are with respect to one variable, holding the other (independent) variables fixed.  
Equation (\ref{eq:211}) is not a unique solution for the exchange current; transverse components not constrained by the two-body WT identity can (and will) be present, but up to these transverse components, (\ref{eq:211}) gives the solution for the exchange current in the limit $q\to0$.

\subsection{Deuteron charge and wave function normalization} \label{sec:charge}

Now, using  the the expression (\ref{eq:211}) for the  two-body interaction current, and the results of Secs.~\ref{subs:E} and \ref{subs:F},  the general expression (\ref{eq:23}) can be evaluated in the limit as $q\to0$.  Expanding the numerators of the (B) diagrams about ${\bf q}=0$, using  (\ref{eq:Bpmterms3}) and the relation
\bea
E_\pm-E_k=\pm\frac{{\bf k}\cdot {\bf q}}{2E_k}+{\cal O}({\bf q}^2) ,
\eea
the deuteron charge becomes (with sums over repeated indices implied)
\begin{widetext}
\begin{align}
2m_d\,&G_C(0)=\lim_{q\to0}J^0_{\lambda\lambda'}(q)=\int_k 
 \overline{\it\Psi}^{\lambda}_{\lambda_n\alpha}(k,P)\,j^0_{\alpha\alpha'}(p,p)\,{\it\Psi}^{\lambda'}_{\alpha'\lambda_n}(k,P)
\nonumber\\
&- e_0\int_k\int_{k'}
 \overline{\it\Psi}^{\lambda}_{\lambda_n\alpha}(k,P)\,h(p)
\Bigg\{\Big(
2 \frac{\partial}{\partial P_0}+\frac{\partial}{\partial k_0}+ \frac{\partial}{\partial k'_0}\Big)  \widetilde V_{\lambda_n\lambda_n',\alpha\alpha'}(k,k';P)\Bigg\}\,h(p')
{\it \Psi}^{\lambda'}_{\alpha'\lambda_n'}(k',P)
\nonumber\\
&-  \int_{k}\frac\partial{\partial k_0}\Bigg\{\frac{m}{k_0}\, \overline{\widetilde{{\Gamma}}}_{\beta\alpha}^{\lambda}(\widetilde k,P)\,\big[h^2(\widetilde  p)\,S(\widetilde p)\big]_{\alpha\alpha'}\,\widetilde{{\Gamma}}_{\alpha'\beta'}^{\lambda'}(\widetilde k,P)\Big[\Lambda(-\widetilde k)\,j^0_0(0)\,\Lambda(-\widetilde k)\Big]_{\beta\beta'}
\Bigg\} \Bigg|_{k_0=E_k}, \label{eq:212}
\end{align}
 where we anticipated the definition of the deuteron charge form factor, $G_C$, previously defined \cite{Gilman:2001yh} and reviewed in Ref.\ II.  
The last term is written in terms of the reduced vertex functions, $\widetilde{\Gamma}$, defined in Eq.~(\ref{eq:reducedGs}).  
Because of the cancellation discussed above, only the strong form factors for particle 2, $h(\widetilde{p})$, remain in  (\ref{eq:212}).  
 
The evaluation of the last term in (\ref{eq:212}) is carried out in Appendix \ref{app:A}.  Combining the result (\ref{eq:312}) with the other terms in Eq.~(\ref{eq:212}) shows that the $k$ and $k'$ derivatives cancel, and the first term is doubled, giving
\bea
2m_d \,e_d=\lim_{q\to0}J^0_{\lambda\lambda'}(q)&=&2 \int_k \overline{\it\Psi}^{\lambda}_{\lambda_n\alpha} (k,P)\,j^0_{\alpha\alpha'}(p,p)\,{\it\Psi}^{\lambda'}_{\alpha'\lambda_n}(k,P)
\nonumber\\
&& -2e_0 \int_k\int_{k'}
 \overline{\it\Psi}^{\lambda}_{\lambda_n\alpha}(k,P)\,h(p)
 \frac{\partial}{\partial P_0}\Big[\widetilde V_{\lambda_n\lambda_n',\alpha\alpha'}(k,k';P)\Big] h(p') \,{\it\Psi}^{\lambda'}_{\alpha'\lambda_n'}(k',P)\, , \label{eq:315}
\eea
\end{widetext}
where $G_C(0)=e_d=1$ is the deuteron charge. In Appendix \ref{app:norm}  it is shown  that  the contributions coming from the $h'$ terms included in the $f_{00}$ and $g_{00}$ contributions to $j^0$ are exactly equal to the $h'$ terms coming from the derivative of the full kenel $\partial \overline{V}/\partial P_0$.  Symbolically, 
\bea
\frac{\partial \overline{V}}{\partial P_0}= \left< j^0\right>\big|_{h'\;{\rm terms}}+\frac{\partial \widetilde V}{\partial P_0} \label{eq:equiv}
\eea
Using $2e_0=1$, this shows that the condition (\ref {eq:315}) is  the same result that emerges from the normalization condition given in Eq.~(2.28) of Ref.~\cite{Adam:1997rb}.

The principal conclusion of this discussion will be summarized for later reference:

(i) The singular diagrams (B$_\pm$) contribute an equal amount to the ``leading'' term [the first term in (\ref{eq:315})].  This is the origin of the definition of the RIA  as equal to {\it two\/}  times the leading diagram (A).

(ii) The $k$ and $k'$ derivatives arising from the (B) diagrams and the interaction current cancel.  This cancellation 
anticipates some general features of the contributions from the interaction current to be discussed in the next Section.

(iii) Because of the relation (\ref{eq:equiv}), the normalization condition can either be written using the bare current ($e_0\gamma^0$) and the derivative of the full kernel, or the dressed current and the derivative of the reduced kernel.  To compute the charge from the dressed current and the derivative of the full kernel introduces an error by double counting the $h'$ terms.

A final comment: including only one factor of $h$ in the definition of the wave function is convenient; it leaves the normalization condition with $h^2j^0\simeq e_0 \gamma^0$ as the leading term.  This one factor occurs  naturally if the vertex function is calculated from the full kernel and the wave function defined  by multiplying by the {\it bare\/} propagator, as was done in Eqs.~(\ref{eq:27}) and (\ref{eq:27a}).

\section{Isoscalar interaction current} \label{sec:ec}

\subsection{General considerations} \label{sec:EX1}

The isoscalar interaction currents are constrained by the two-body WT identity (\ref{eq:29}) [or, alternatively, (\ref{eq:210})].  If the right-hand-side of this identity is zero, then the longitudinal component of the current is zero, and the assumption that the accompanying transverse component is also zero is the simplest possible choice for the interaction current.   Only  if the longitudinal component of the current is non-zero will we look to the physics to  see if a non-zero transverse component should accompany the longitudinal component.   We will refer to this assumption as the {\it principal of simplicity\/}. 

A second principal that governs our choice of interaction current is the {\it principal of picture independence\/}.  This principal leads to very strong constraints that all but uniquely define the interaction currents, at least for isoscalar interactions.  This principal will be developed in detail in the latter part of this section.

\begin{figure}
\centerline{
\mbox{
\includegraphics[width=3.1in]{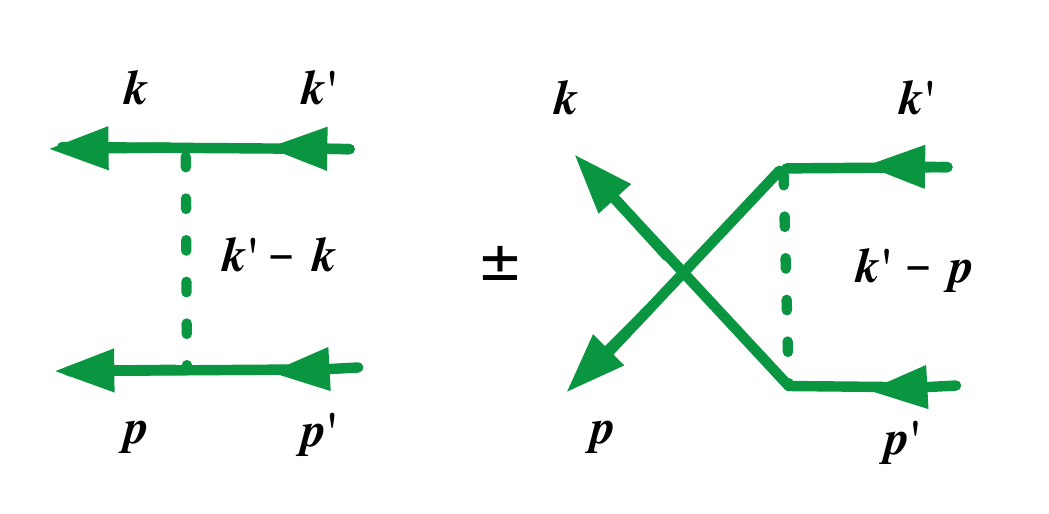}
}
}
\caption{\footnotesize\baselineskip=10pt (Color on line) Generic Feynman meson exchange diagrams for the direct (left panel) and exchange (right panel) processes, showing the two momentum transfers $q_d=k'-k$ and $q_e=k'-p=k'+k-P$.  }
\label{Fig4}
\end{figure} 

In the CST, the kernel (or reduced kernel) is symmetrized (or antisymmetrized) by hand, so it is the sum of two terms
\bea
\widetilde{V}_{\beta\beta',\alpha\alpha'}(k,k';P)=\sfrac12\Big[&&V_{\beta\beta',\alpha\alpha'}(k,k';P)
\nonumber\\
&&\pm V_{\alpha\beta',\beta\alpha'}(P-k,k';P)\Big],\qquad \label{eq:4.1a}
\eea
where $\{k,\beta; p\equiv P-k,\alpha\}$ ($\{k',\beta'; p'\equiv P-k',\alpha'\}$) are the outgoing (incoming) four-momenta and Dirac indices of the two particles, $\widetilde{V}$ is the reduced kernel, and we emphasize that $V$ (without the bar) is the {\it unsymmetrized\/} reduced kernel (this is similar to Eq.~(2.6) of Ref.~\cite{Gross:2008ps}, but here the strong nucleon form factors have been removed and $k$ is the momentum of the on-shell particle 1, instead of the relative momentum).  The linear combination (\ref{eq:4.1a}) is symmetric or antisymmetic under the exchange of the particles in the final state: $\{k,\beta\}\leftrightarrow\{P-k,\alpha\}$.  The kernels used in the CST are linear combinations of such terms, as discussed in detail in Ref.~\cite{arXiv:1007.0778}.  For the one boson exchange (OBE) models being discussed in this paper, the Feynman diagrams corresponding to the two terms are shown in Fig.~\ref{Fig4}.

\begin{figure}
\leftline{
\mbox{
\includegraphics[width=3.7in]{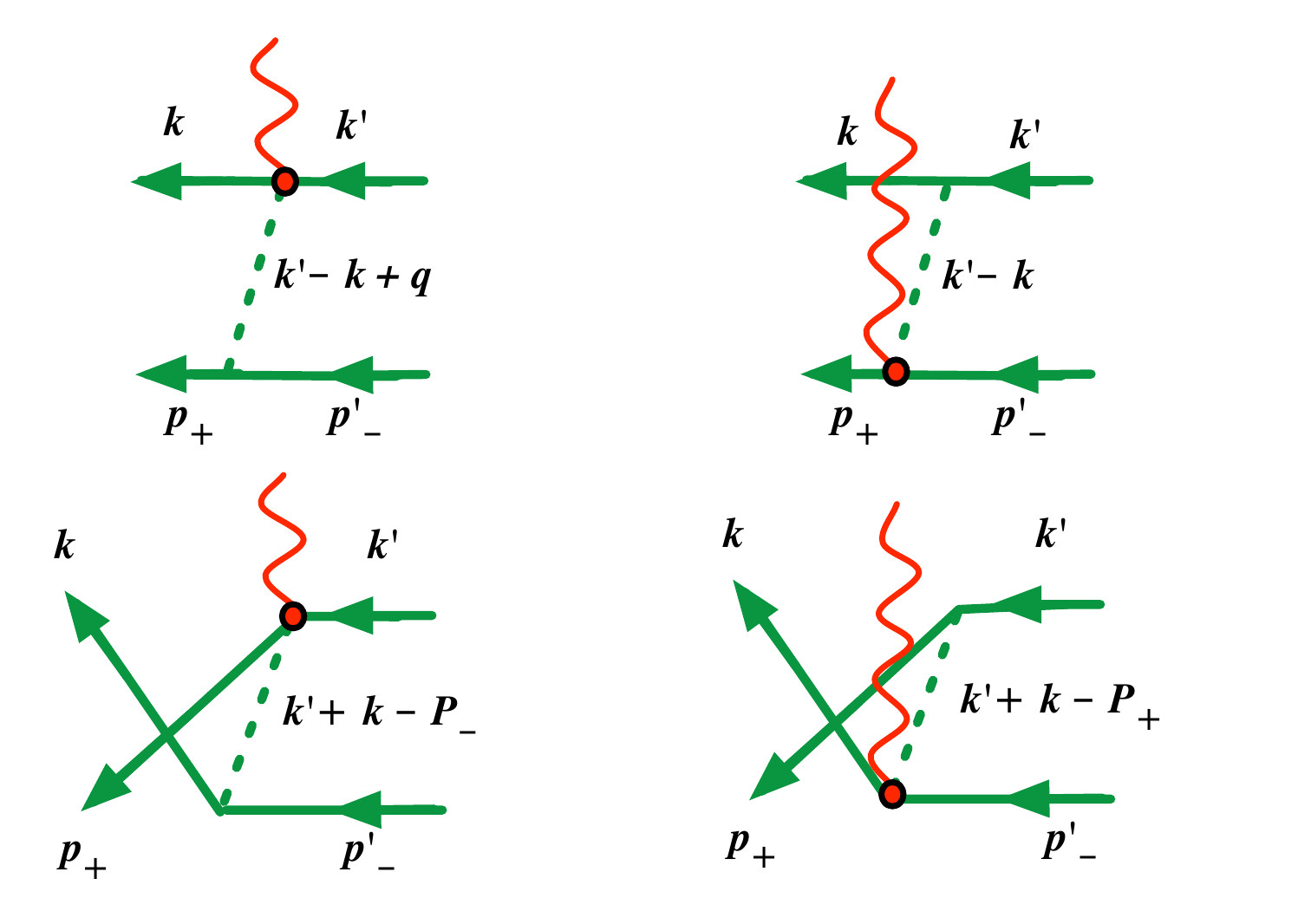}
}
}
\caption{\footnotesize\baselineskip=10pt (Color on line) Feynman diagrams for the generic meson interaction currents arising from vertex interactions.  Top row: direct terms;  bottom row: exchange terms.  }
\label{Fig3}
\end{figure} 

It is shown in Appendix \ref{app:intercur} that any meson exchange interaction that depends {\it only\/} on the exchanged four-momentum will not contribute to the right hand side of the two-body WT identity (\ref{eq:210}), and it can therefore be assumed, guided by the principal of simplicity, that the interaction current accompanying this interaction is also zero.

Therefore, isoscalar interaction currents will only come from a possible energy dependence of the vertex functions of the OBE interaction.  These generate  contact interactions of the type shown in  Fig.~\ref{Fig3}.  Examination of the vertex functions used in models WJC-1 and WJC-2 \cite{Gross:2008ps,arXiv:1007.0778} shows the this energy dependence is only located in the off-shell couplings, i.e. the terms proportional to $\Theta(p)$ [where $\Theta$ was defined in Eq.\ (\ref{eq:theta})].  These terms do not exist in theories where the particles are always on-shell, and are unique to the CST.  Such $\Theta$ dependent terms are present in the scalar, vector, and pseudoscalar exchange terms, but the $\Theta$ terms in the pseudoscalar exchange reduce to terms depending on momentum transfer only (and hence their contributions to isoscalar exchange currents cancel).

\subsection{General structure of the currents}

This discussion begins by using minimal substitution to find the general form of the interaction currents.  Then, in a subsequent section, it is shown how the principal of picture independence leads a unique choice for these currents.

The unsymmetrized reduced OBE kernel introduced in (\ref{eq:4.1a}) has the form
\bea
{V}_{\beta\beta',\alpha\alpha'}(k,p;k',p')=\sum_b V_{\beta\beta',\alpha\alpha'}^b(k,p;k',p')
\qquad \label{eq:OBE}
\eea
where 
$b$ is the boson type, and it will be convenient in this section to use the redundant notation $\{k,p;k',p'\}$ in place of $\{k,k';P\}$.    
When the external momenta are labeled as above this is referred to as the {\it direct\/} kernel; when $\{k,\beta\}$ and $\{p,\alpha\}$ are exchanged it will be referred to as the {\it exchange\/} kernel.  Each term in the sum has the form
\bea
V_{\beta\beta',\alpha\alpha'}^b(k,p;k',p')&=&\Lambda^b_{\beta\beta'}(k,k')\otimes\Lambda^b_{\alpha\alpha'}(p,p') 
\widetilde\Delta^b(\tilde q) \qquad\quad 
\nonumber\\
&=&\mathring{V}^{b}_{\beta,\beta'}(k,k')\otimes \Lambda^b_{\alpha\alpha'}(p,p') 
\nonumber\\
&=&\Lambda^b_{\beta\beta'}(k,k') \otimes \mathring{V}^{b}_{\alpha,\alpha'}(p,p') 
 \label{eq:ONEkernel}
\eea
where $\widetilde\Delta^b(\tilde q)$, with $\tilde q=k'-k=p-p'$, 
is the dressed meson propagator, including the meson form factor and some additional factors 
\bea
\widetilde\Delta^b(\tilde q)&=&\epsilon_b\,\delta\, \frac{f(\Lambda_b,\tilde q)}{m_b^2+|\tilde q^2|}=\epsilon_b\,\delta\,\Delta^b(\tilde q) \, ,\label{eq:4.10a}
\eea
and $\Lambda^b_{\beta\beta'}(k,k')$ or $\Lambda^b_{\alpha\alpha'}(p,p')$ are the $bNN$ vertex functions for particle $1$ or $2$ \cite{Gross:2008ps}.   These vertex functions should not be confused with the projection operators $\Lambda(k)$ introduced in Eq.~(\ref{eq:spindecom}).   The second and third lines of (\ref{eq:ONEkernel}) define {\it truncated OBE\/} kernels (that part of the kernel that remains once the $bNN$ vertex function for particle 1 or 2 has been removed).  These will be used in the discussion below.  The vertex functions, $\Lambda^b_{\alpha\alpha'}(p,p')$ are decomposed into two parts 
\bea
\Lambda^b_{\alpha\alpha'}(p,p')&=&A^b_{\alpha\alpha'}(p-p')
\nonumber\\&&
+\big[B^b\Theta(p')+\Theta(p)B^b\big]_{\alpha\alpha'} \qquad \label{eq:4.10b}
\eea
where, for each meson
\bea
A^b(p-p')=\begin{cases} 
g_s{\bf 1} & {\rm scalar} \cr
g_p\gamma^5 & {\rm pseudoscalar}  \cr
g_v\,O^\mu_v(p-p')& {\rm vector}\cr
g_a\gamma^5\gamma^\mu & {\rm pseudovector} \end{cases}\qquad\quad \label{eq:Aterms}
\eea
%
with $O^\mu_v(p-p')=\gamma^\mu+\kappa_v\,i\sigma^{\mu\nu}(p-p')_\nu/(2m)$, and
\bea
B^b=\begin{cases} 
-\nu_s{\bf 1} & {\rm scalar} \cr
-g_p(1-\lambda_p)\gamma^5 & {\rm pseudoscalar}  \cr
\phantom{-}g_v\nu_v\,\gamma^\mu& {\rm vector}\cr
\phantom{-}0 & {\rm pseudovector}\, . \end{cases} \label{eq:Bs}
\eea
Note that the $B$'s are independent of momenta, and the $A$'s depend only on $\pm\tilde q$ (there is also another dependence on $\tilde q$ coming from the vector meson propagator, but this does not affect the discussion).  With the proper substitutions, the same relations hold for $\Lambda^b_{\beta\beta'}(k,k')$, and for the exchange kernel.

\begin{figure*}
\centerline{
\mbox{
\includegraphics[width=5in]{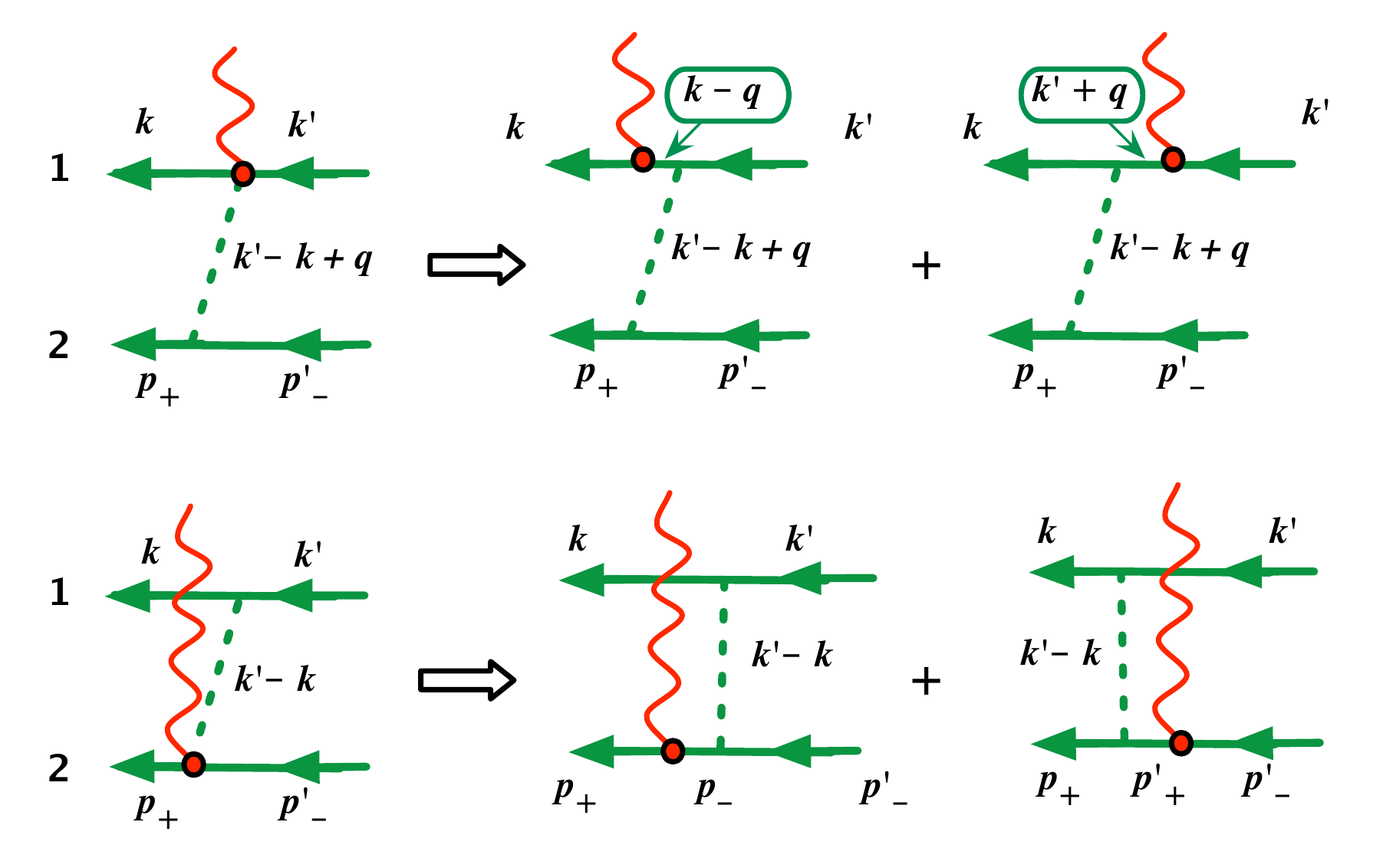}
}
}
\caption{\footnotesize\baselineskip=10pt (Color on line) Illustration of the structure of the direct interaction currents displayed in Eq.~(\ref{eq:4.7}).  In the diagram the meson currents $j_b^\mu$ are attached to each of the four nucleon lines, but will not be identified as nucleon currents until later. }
\label{Figx-d}
\end{figure*} 
\begin{figure*}
\centerline{
\mbox{
\includegraphics[width=6in]{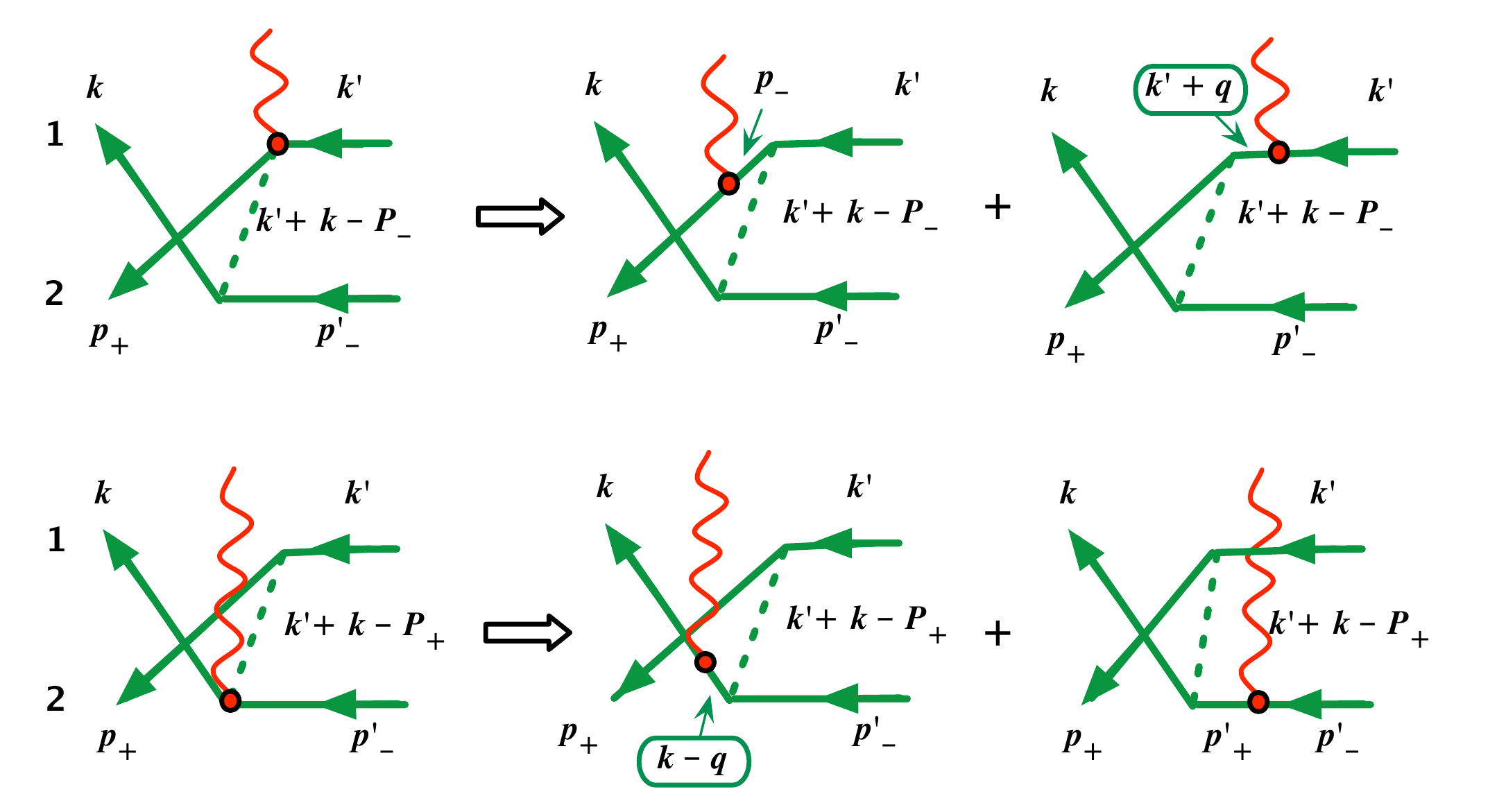}
}
}
\caption{\footnotesize\baselineskip=10pt (Color on line) Illustration of the structure of the exchange interaction currents displayed in Eq.~(\ref{eq:4.7}).  The meson currents $j_b^\mu$ attached to the outgoing crossed nucleon lines to are identified with the number of the  particle in the final state, so that (for example) the meson current in the lower right diagram is $j_{b1}^\mu$ because the outgoing nucleon will become particle 1, even though $j_b^\mu$ attaches to particle 2 at the vertex. }
\label{Figx-e}
\end{figure*} 

For any particle with momentum $p$,  minimal substitution leads to the replacement
\bea
p^\mu\to p^\mu-e_0 A^\mu\, ,
\eea
where $e_0$ is the isoscalar nucleon charge and commutes with all isospin operators.  This means that, in general, there will be contributions from both isovector and isoscalar meson exchanges.  However, the factors in pseudoscalar vertex functions depend only on the momentum transfer $\tilde q$, and will not contribute to the current because the $p$ and $p'$ terms (or the $k$ and $k'$ terms) will cancel.  
Nevertheless, it is convenient to ignore this now (so that all the mesons can be treated on the same footing) and simply drop contributions from pseudoscalar mesons ($\pi$ and $\eta$) in the final result.  

Since the projection operators $\Theta$ are linear in $p$, minimal substitution gives
\bea
\Theta(p_i)\to \frac{e_0}{2m}\,\gamma^\mu\to \frac{1}{2m}\,j_{b}^\mu(q) \, , \label{eq: thetacurrent}
\eea
where $j_b^\mu$ is the generalization of the point-like interaction current,   $e_0\gamma^\mu$, which might arise from the $bNN$ vertex.  The specific structure of $j_b^\mu$ will be fixed later.  At this point, conditions on $j_b^\mu$ will be determined by demanding that the interaction current satisfy the two-body WT identity.
The effect of the replacement (\ref{eq: thetacurrent}) is to construct the interactions currents from the kernels by replacing, for example, the vertex function $\Lambda^b(p,p')$ by meson current operators using the following substitution
\bea
\Lambda^b(p,p')\to \frac1{2m}\left[B^b\,j^\mu_b + j^\mu_b\,B^b\right]\, .
\eea
In terms of the truncated OBE kernels introduced in (\ref{eq:ONEkernel}), this replacement gives, for the contribution coming from vertex function $\Lambda^b(p,p')$,
\bea
&&V_{\beta\beta',\alpha\alpha'}^{b\,\mu}(k,p_+;k',p'_-)
\nonumber\\
&&\qquad=\frac1{2m}\mathring{V}^{b}_{\beta,\beta'}(k,k')
[B^bj_b^\mu+j_b^\mu B^b]_{\alpha\alpha'}(p_+,p'_-),\qquad\quad
\eea
where the arguments of the current used originally in Eq.~(\ref{eq:VandVmu}), $\{kP_+;k'P_-\}$, are, for convenience, replaced in this section by the momenta of the individual particles $\{k,p_+;k',p'_-\}$, with $p_-'=P_--k'$ and $p_+=P_+-k$.  The labeling of the momenta is illustrated in the bottom row of  Fig.~\ref{Figx-d}.

\begin{figure*}[t]
\centerline{
\mbox{
\includegraphics[width=5in]{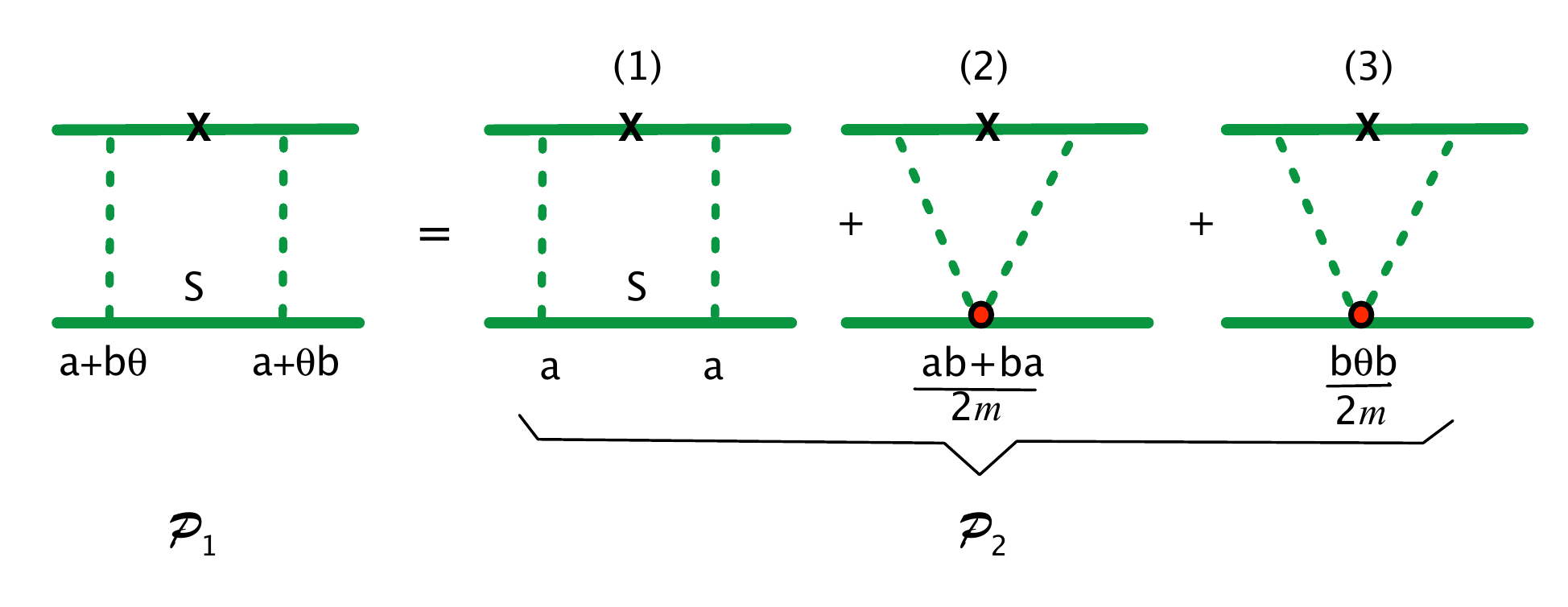}
}
}
\caption{\footnotesize\baselineskip=10pt (Color on line) Two equivalent pictures of the generic meson interaction up to fourth order.  In picture ${\cal P}_1$ the fourth order contribution is a single box diagram with off shell-couplings represented by the operator $\Theta$.  In picture ${\cal P}_2$ the single box is  rearranged into the three Feynman diagrams that emerge when the cancellations between $\Theta$ and the nucleon propagator $S$ are exploited using $\Theta S=1/(2m)$.   Here  
$a\simeq hAh'$ and $b\simeq hBh'$ with $A$ and $B$ given in Eq.~(\ref{eq:4.10b}).
}
\label{fig:P1&P2}
\end{figure*} 

The exchange terms will have the appropriate indices and momenta interchanged.  Using these truncated OBE kernels and labeling the momenta of the various terms in the interaction current as shown in Figs.~\ref{Figx-d} and \ref{Figx-e} leads to the following ansatz for the  reduced interaction current: 
\begin{widetext}
\bea
\widetilde{V}_{\beta\beta',\alpha\alpha'}^{\mu}(k,p_+;k',p'_-)=&&\frac{1}{4m}\sum_b \Big\{[B^b\,j_b^{\mu}+j_b^\mu\,B^b]_{\beta\beta'}(k,k') \otimes \mathring{V}^{b}_{\alpha\alpha'}(p_+,p'_-)+
\mathring{V}^{b}_{\beta\beta'}(k,k')\otimes [B^b\,j_b^{\mu}+j_b^\mu\,B^b]_{\alpha\alpha'}(p_+,p'_-)
\nonumber\\
&&\pm[B^b\,j_b^{\mu}+j_b^\mu\,B^b]_{\alpha\beta'}(p_+,k') \otimes \mathring{V}^{b}_{\beta\alpha'}(k,p'_-)
\pm\mathring{V}^{b}_{\alpha\beta'}(p_+,k')\otimes [B^b\,j_b^{\mu}+j_b^\mu\,B^b]_{\beta\alpha'}(k,p'_-)\Big\} \, .\qquad\quad
\label{eq:4.7}
\eea
where the extra factor of 1/2 comes from the symmetrization, Eq.~(\ref{eq:4.1a}).  Note that the  four terms in the first line come from the direct interaction, with moments labeled as in Fig.~\ref{Figx-d} [with the top row corresponding to the first two terms dependent on $\mathring{V}^{b}(p_+,p'_-)$ and the bottom row of the figure corresponding to the last two terms dependent of $\mathring{V}^{b}(k,k')$].  The four terms in the second line of Eq.~(\ref{eq:4.7}) come from the exchange interaction, with momenta labeled as in Fig.~\ref{Figx-e} [with the top row of the figure corresponding to the first two terms dependent of $\mathring{V}^{b}(k,p'_-)$ and the bottom row corresponding to the last two terms dependent on $\mathring{V}^{b}(p_+,k')$].   Using the form (\ref{eq:210}) for the WT identity (and expanding the symmetrized kernels in terms of unsymmetrized kernels) and writing the unsymmetrized kernels in terms of $\mathring{V}^{b}$ 
leads to the requirement
\bea
q_\mu&&\widetilde{V}_{\beta\beta',\alpha\alpha'}^{\mu}(k,p_+;k',p'_-)
\nonumber\\
&&=\frac{1}{4m}\sum_b \Big\{[B^b\,q_\mu j_b^{\mu}+q_\mu j_b^\mu\,B^b]_{\beta\beta'}(k,k') \otimes \mathring{V}^{b}_{\alpha\alpha'}(p_+,p'_-)+\mathring{V}^{b}_{\beta\beta'}(k,k')\otimes [B^b\,q_\mu j_b^{\mu}+q_\mu j_b^\mu\,B^b]_{\alpha\alpha'}(p_+,p'_-)
\nonumber\\
&&\qquad\qquad\pm[B^b\,q_\mu j_b^{\mu}+q_\mu j_b^\mu\,B^b]_{\alpha\beta'}(p_+,k') \otimes \mathring{V}^{b}_{\beta\alpha'}(k,p'_-)
\pm\mathring{V}^{b}_{\alpha\beta'}(p_+,k')\otimes [B^b\,q_\mu j_b^{\mu}+q_\mu j_b^\mu\,B^b]_{\beta\alpha'}(k,p'_-)\Big\} 
\nonumber\\
&&=\frac{e_0}{2}\Big\{V_{\beta\beta',\alpha\alpha'}(k,p_-;k',p'_-)-V_{\beta\beta',\alpha\alpha'}(k,p_+;k',p'_+) 
+V_{\beta\beta',\alpha\alpha'}(k-q,p_+;k',p_-')-V_{\beta\beta',\alpha\alpha'}(k,p_+;k'+q,p_-')
\nonumber\\
&&\qquad+V_{\alpha\beta',\beta\alpha'}(p_-,k;k',p'_-)-V_{\alpha\beta',\beta\alpha'}(p_+,k;k',p'_+) 
+V_{\alpha\beta',\beta\alpha'}(p_+,k-q;k',p_-')
-V_{\alpha\beta',\beta\alpha'}(p_+,k;k'+q,p_-')\Big\}\nonumber\\
&&=\frac{e_0}{2}\sum_b\Big\{\mathring{V}^{b}_{\beta\beta'}(k,k')\otimes [\Lambda^b(p_-,p_-')-\Lambda^b(p_+,p_+')]_{\alpha\alpha'} 
+[\Lambda^b(k-q,k')-\Lambda^b(k,k'+q)]_{\beta\beta'}\otimes\mathring{V}^{b}_{\alpha\alpha'}(p_+,p_-')
\nonumber\\
&&\qquad\qquad\pm\mathring{V}^{b}_{\alpha\beta'}(p_+,k')\otimes [\Lambda^b(k-q,p_-')-\Lambda^b(k,p_+')] _{\beta\alpha'}
\pm[\Lambda^b(p_-,k')-\Lambda^b(p_+,k'+q)] _{\alpha\beta'}\otimes\mathring{V}^{b}_{\beta\alpha'}(k,p_-')\Big\},\qquad\quad
\label{eq:4.7WT}
\eea
\end{widetext}
where the terms in the second step must be rearranged in order to display them in terms of the differences shown in the last step.   Using the general form (\ref{eq:4.10b}) for the $\Lambda$'s, and recalling that $A^b$ factor contained in the vertex function $\Lambda^b(p,p')$ [recall Eq.~(\ref{eq:Aterms})] only depends on $p-p'$,  it follows that the differences, when expressed as a matrix in Dirac space all reduce to the same result
\bea
\frac1{2m}\Big[B^b\slashed{q}+\slashed{q}B^b\Big]&&=\Lambda^b(p_-,p_-')-\Lambda^b(p_+,p'_+)
\nonumber\\&&
=\Lambda^b(k-q,k')-\Lambda^b(k,k'+q)
\nonumber\\&&
=\Lambda^b(k-q,p_-')-\Lambda^b(k,p_+')
\nonumber\\&&
=\Lambda^b(p_-,k')-\Lambda^b(p_+,k'+q)\, .\qquad
\eea
This in turn shows that the WT identity (\ref{eq:4.7WT}) is satisfied if all the boson currents satisfy the same simple condition 
\bea
q_\mu\,j_{b}^\mu=&&e_0\,\slashed{q}\,  . \label{eq:WTforjb}
\eea
%

\begin{figure*}[t]
\centerline{
\mbox{
\includegraphics[width=5in]{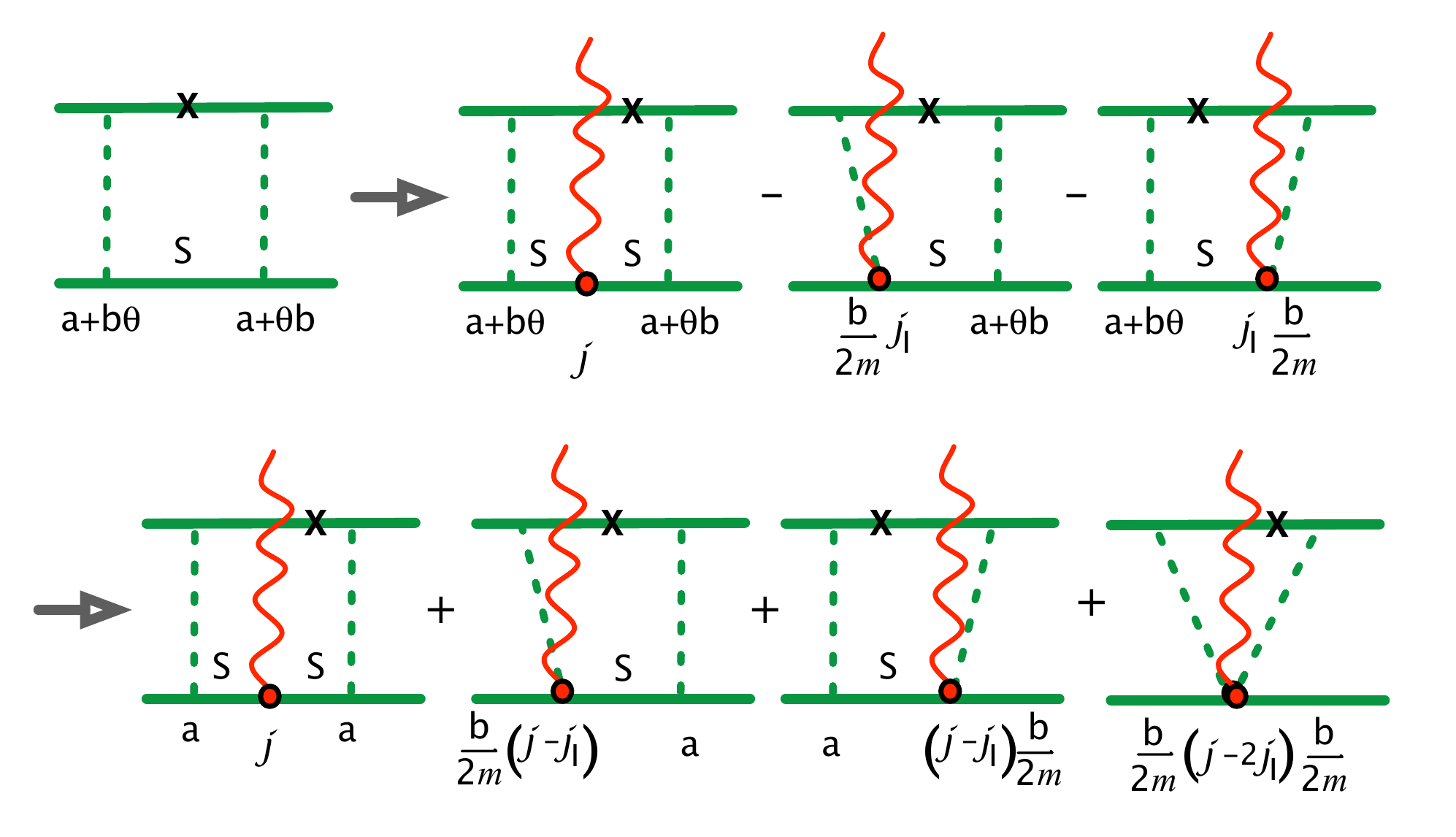}
}
}
\caption{\footnotesize\baselineskip=10pt (Color on line) The single box diagram for picture ${\cal P}_1$ generates, through a two-step process, four diagrams with the current coupling to the off-shell particle 2.  Here $h=h'=1$ and $j\simeq j_0$ defined in Eq.~(\ref{eq:j0current}) is the  nucleon current, and $j_I\simeq j_b$, where $j_b$ was introduced in Eq.~(\ref{eq: thetacurrent}). }
\label{fig:P1}
\end{figure*} 

\begin{figure*}
\centerline{
\mbox{
\includegraphics[width=6.5in]{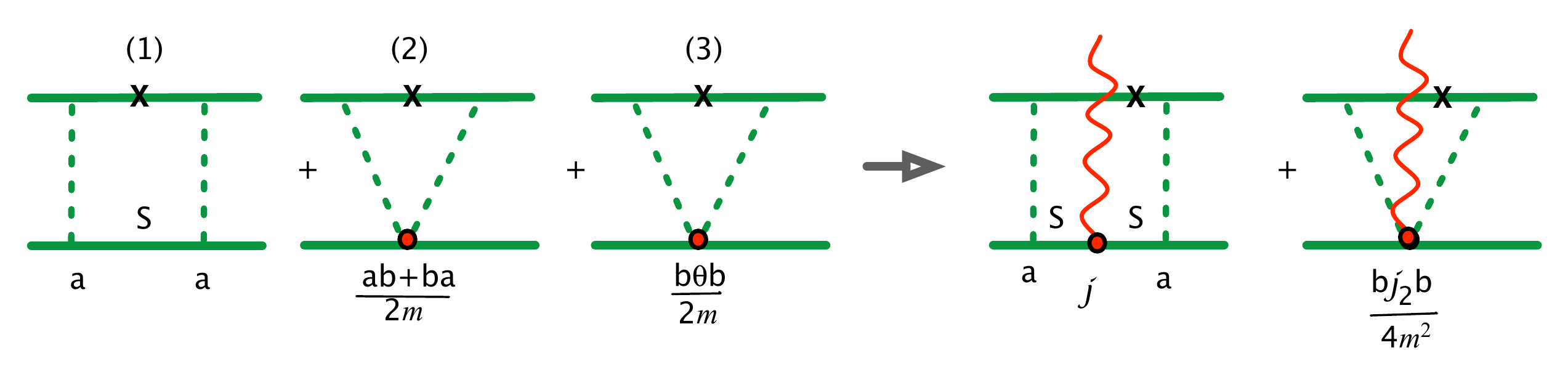}
}
}
\caption{\footnotesize\baselineskip=10pt (Color on line) The three diagrams for picture ${\cal P}_2$ generate only two diagrams with the current coupling to particle 2. }
\label{fig:P2}
\end{figure*} 

\begin{figure*}
\centerline{
\mbox{
\includegraphics[width=7in]{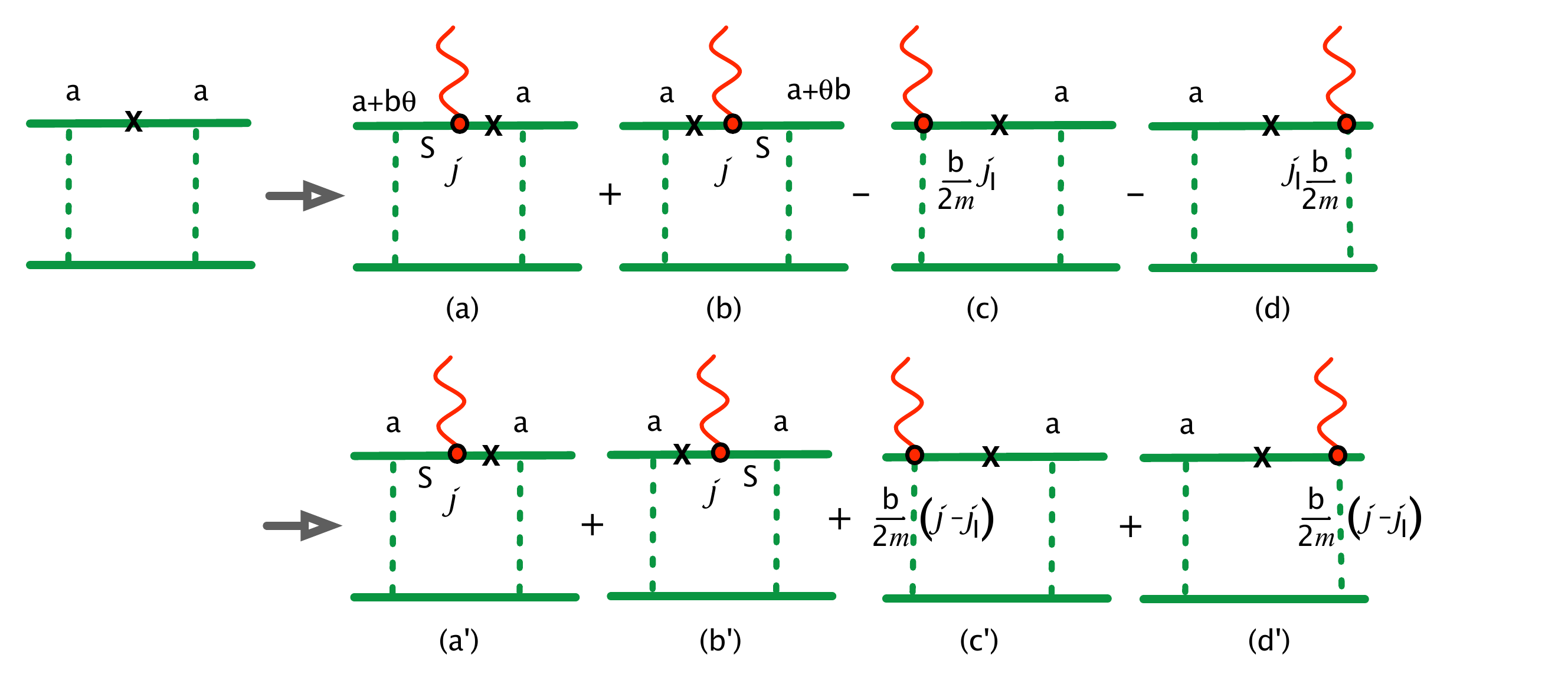}
}
}
\caption{\footnotesize\baselineskip=10pt (Color on line) The single box diagram for picture ${\cal P}_1$ generates, again through a two-step process, four diagrams with the current coupling to the on-shell particle 1.  With the constraint (\ref{eq:picconstraint}) only the two diagrams depending on $a^2$ survive, giving the same result as picture ${\cal P}_2$.}
\label{fig:P1-1}
\end{figure*} 

While this condition is satisfied by the simple ansatz, $j_{b}^\mu= e_0\,\gamma^\mu$, this is not a satisfactory choice because it has a point-like structure inconsistent with the extended structure of the nucleon and the mesons being exchanged.  There should be an electromagnetic form factor associated with the $\gamma bNN$ vertex.  Following the treatment of Ref.~\cite{Gross:1987bu} this form factor can be incorporated into a transverse part of the current.  One solution is
\bea
j^\mu_{b}(q)=e_0\,\Big\{(F_{\gamma b NN}(Q^2)-1)\tilde\gamma^\mu+\gamma^\mu\Big\}\, ,\qquad \label{eq:4.13}
\eea
where $F_{\gamma bNN}(0)=1$ and the transverse $\tilde{\gamma}^\mu$ was defined in Eq.~(\ref{eq:gammatilde}).
Note that this current is finite as $q^2\to0$.  Furthermore, when contracted with another conserved current (or a photon polarization vector) the term proportional to $q^\mu$ can be dropped, giving 
\bea
j_{b}^\mu(q)\to e_0\,F_{\gamma bNN}(Q^2)\,\gamma^\mu\, .
\label{eq:4.14}
\eea
The interaction currents (\ref{eq:4.13}) are very general and provide very little predictive power.  It will now be shown how these currents can be fixed uniquely by the principle of picture independence.

\subsection{Constraints imposed by the principal of picture independence} \label{sec:rules}

 Previous  studies of three-body forces in the CST \cite{Stadler:1996ut,Gross:2008ps} have shown how the off-shell couplings arising from the dependence of the vertex functions on the projection operator $\Theta(p)$ can be removed if their effect is reproduced by adding other interactions.  This leads to two equivalent {\it pictures\/}, to be denoted in this paper by ${\cal P}_1$ and ${\cal P}_2$.  Picture ${\cal P}_1$ is the original  pure OBE interaction model with off-shell couplings.  Picture ${\cal P}_2$ is a {\it dynamically equivalent\/} model with {\it no\/} off-shell couplings, but with additional  interactions added to reproduce the effect of the off-shell couplings.    The same two pictures can be described in the electromagnetic interactions of two-body systems, and requiring that they give an identical description of the physics leads to strong constraints on the details of the interaction currents in both pictures.

Begin the discussion by looking at the two pictures up to fourth order, illustrated in Fig.~\ref{fig:P1&P2}.  The action of the off-shell projection operator $\Theta$ on the nucleon propagator $S$ removes the internal nucleon line, shrinking neighboring interactions to a point and replacing some of the box diagrams by triangle diagrams.  A simple box diagram in picture ${\cal P}_1$ is converted into three diagrams in  ${\cal P}_2$, a box (without off-shell couplings) and two triangles.   To construct a conserved current using the methods of Ref.~\cite{Gross:1987bu}, the kernel must be of the form $h(p)\tilde Vh(p')$  (with $\tilde V$  independent of $h$), and the triangle diagrams  that emerge from picture ${\cal P}_2$ are proportional to four powers of $h$. Therefore, only if $h=1$ is it easy to construct the exact current operator in both pictures.   
(It may be possible to construct the current in picture ${\cal P}_2$ for the case $h\ne1$, but this has not been investigated, and it is not necessary to do so.)  As it turns out, this limitation does not seem to be serious; it will be shown that the current constructed for $h=1$ is also an acceptable choice for the more general case when $h\ne1$.

Each of these two equivalent sets of diagrams suggests its own current operator.  Limiting the discussion to the current of off-shell particle 2, the current diagrams  generated by picture ${\cal P}_1$ are shown in Fig.~\ref{fig:P1}, where because of the restriction $h=1$ (for this argument only) the nucleon current $j^\mu\to j_0^\mu$, where $j_0^\mu$ was defined in Eq.~(\ref{eq:j0current}).  The single Feynman box diagram generates four current diagrams, all of which include contributions from the nucleon current, $j_0^\mu$ and the interaction current $j_{b}^\mu$ generated from the off-shell $\Theta$ terms in the $bNN$ interaction.  The current diagrams generated by picture ${\cal P}_2$ are shown in Fig.~\ref{fig:P2}. Here there are only two diagrams, since one of the triangle diagrams, Fig.~\ref{fig:P2}(2), does not have a coupling depending on the triangular loop momentum, while the other, Fig.~\ref{fig:P2}(3), generates a new current associated with the momentum dependent part of the $bbNN$ interaction.  

The constraint arising from the principal of picture independence  requires that the currents appearing in the two pictures be identical.  Since ${\cal P}_2$ has no $bNN$ interaction current, that must also vanish for ${\cal P}_1$, and this will happen only if %
\bea
j_b^\mu(q)=j_0^\mu(q)\, . \label{eq:picconstraint}
\eea
This requirement also fixes the $bbNN$ current in ${\cal P}_2$, and the (infinitely) many other higher order interaction currents that are a feature of ${\cal P}_2$.   
Fortunately,  these higher order currents need not be calculated when using picture ${\cal P}_1$; they all are a consequence of the simple OBE interaction current.   There is an equivalence theorem similar to that found in previous studies of the three body  system: {\it the simple interaction currents arising from a theory with off-shell couplings are equivalent to an infinite number of very complex interaction currents that arise from a theory with no off-shell couplings\/}.  Further discussion of this fascinating subject is postponed for a later day.

Note that the constraint (\ref{eq:picconstraint}) is possible only because both currents satisfy the same conditions, (\ref{eq:WTforj0}) for $j_0^\mu$ and (\ref{eq:WTforjb}) for $j_b^\mu$.
As anticipated, the hadronic structure of the currents from {\it all\/} boson exchanges are identical (even though their Dirac structure is different).  Complete equivalence also requires that the magnetic, purely transverse parts of the currents be identical, a result that might not have been anticipated.

Figure~\ref{fig:P1-1} shows that the same conclusions also apply to the electromagnetic interactions of the on-shell particle 1 (those interactions giving rise to the (B) diagrams of Fig.~\ref{Fig1}).  Here picture ${\cal P}_1$ retains off-shell couplings and interactions from those parts of the diagram where particle 1 is off-shell.  Among these are the contributions illustrated  in Fig.~\ref{fig:P1-1}(c) and (d); these are   interaction currents arising from the projection operator in the $bNN$ vertex associated with off-shell contributions of particle 1, even though the  particle entering (or leaving) the interaction is on-shell.  Again, the off-shell projection operator cancels the neighboring nucleon propagator, giving contributions that exactly cancel the interaction current terms 
reducing the total result to only two diagrams, Figs.~\ref{fig:P1-1}(a') and (b'), the same two that would appear in picture ${\cal P}_2$, insuring picture equivalence.  

Examination of Fig.~\ref{fig:P1-1} illustrates another point:  in the framework of picture ${\cal P}_1$, the exact answer for the combined electromagnetic contributions from particle 1 (the (B) diagrams  of Fig.~\ref{Fig1}  together with the interaction currents from particle 1) is the result of a cancellation between the off-shell contributions and the interaction currents.  This cancellation also holds when $h\ne1$, and will be formulated more precisely  in the next section.

\subsection{Computation of interaction current contributions}

The condition (\ref{eq:picconstraint}) permits the interaction current to be re-expressed in terms of four new {\it truncated\/} $NN$ kernels (to be distinguished from the truncated OBE kernels introduced above), which are denoted $V^{\ell \ell'}$ (with $\ell=\{1,2\}$ and $\ell'=\{i,f\}$ as described below) and the nucleon current $j_0^\mu$.  Returning to the notation used in Eq.~(\ref{eq:VandVmu}) for the arguments of the current and the kernel, and noting that $k$ and $k'$ are on shell ($k^2=k'^2=m^2$), the interaction current is
\begin{widetext}
\bea
\widetilde{V}^\mu_{\beta\beta',\alpha\alpha'}(k P_+; k' P_-)&=&\frac1{2m}\Big\{{V}^{2i}_{\beta\beta',\alpha\alpha_1}(k,k';ÊP_+)\,j^\mu_{0\,\alpha_1\alpha'}(q)+j^\mu_{0\,\alpha\alpha_1}(q)\,{V}^{2f}_{\beta\beta',\alpha_1\alpha'}(k, k'; P_-)
\nonumber\\
&&+{V}^{1i}_{\beta\beta_1,\alpha\alpha'}(k,k'+q; P_+)\,j^\mu_{0\,\beta_1\beta'}(q)+j^\mu_{0\,\beta\beta_1}(q)\,{V}^{1f}_{\beta_1\beta',\alpha\alpha'}(k-q,k'; P_-) \Big\}\, ,\label{eq:Icurrent}
\eea
and using the notation $k_f$ and $k_i$ for four-momenta that are not necessarily on-shell, the truncated $NN$ kernels are 
\bea
{V}^{1i}_{\beta\beta',\alpha\alpha'}(k_f,k_i; P)&=&\frac1{2}\sum_b\Big\{B^b_{\beta\beta'}\otimes\mathring{V}^{b}_{\alpha\alpha'}(p_f,p_i)
\pm B^b_{\alpha\beta'}\otimes\mathring{V}^{b}_{\beta\alpha'}(k_f,p_i)\Big\}
\nonumber\\
{V}^{1f}_{\beta\beta',\alpha\alpha'}(k_f,k_i; P)&=&\frac1{2}\sum_b\Big\{B^b_{\beta\beta'}\otimes\mathring{V}^{b}_{\alpha\alpha'}(p_f,p_i)
\pm B^b_{\beta\alpha'}\otimes\mathring{V}^{b}_{\alpha\beta'}(p_f,k_i)\Big\}
\nonumber\\
{V}^{2i}_{\beta\beta',\alpha\alpha'}(k,k';P)&=&\frac1{2}\sum_b\Big\{B^b_{\alpha\alpha'}\otimes\mathring{V}^{b}_{\beta\beta'}(k,k')
\pm B^b_{\beta\alpha'}\otimes\mathring{V}^{b}_{\alpha\beta'}(p,k')\Big\}
\nonumber\\
{V}^{2f}_{\beta\beta',\alpha\alpha'}(k,k';P)&=&\frac1{2}\sum_b\Big\{B^b_{\alpha\alpha'}\otimes\mathring{V}^{b}_{\beta\beta'}(k,k')
\pm B^b_{\alpha\beta'}\otimes\mathring{V}^{b}_{\beta\alpha'}(k,p')\Big\}
\, , \label{eq:reducedk}
\eea
\end{widetext}
with $p_i=P-k_i$ and $p_f=P-k_f$.
Note that the kernel $V^{1i}$ accompanies the current $j_0$ of the  {\it incoming\/} nucleon 1, $V^{1f}$ accompanies the current $j_0$ of the  {\it final\/} nucleon 1,  $V^{2i}$ accompanies the current $j_0$ of the  {\it incoming\/} nucleon 2, and $V^{2f}$ accompanies the current $j_0$ of the  {\it final\/} nucleon 2.  Also observe that, in applications, $V^{1i}$ always has $k_f=k$ (on-shell) and  $V^{1f}$ always has $k_i=k'$ (on-shell), so that in no case are both $k_f$ and $k_i$ off-shell in the same term.

For future reference it is useful to note that $V^{2i}$ is the coefficient of the of the off-shell projection operator $\Theta(p')$, $V^{2f}$ the coefficient of the off-shell projection operator $\Theta(p)$, and, when either $k=k_f$ or $k'=k_i$ are off-shell, $V^{1i}$ the coefficient of the off-shell projection operator $\Theta(k_i)$, and $V^{1f}$ the coefficient of the off-shell projection operator $\Theta(k_f)$.  The exchange term of the full kernel contains a term proportional to $\Theta(p)\otimes \Theta(p')$ so it is incorrect to expand the full kernel in a sum of the form $V^{2f}\otimes\Theta(p) + V^{2i}\otimes\Theta(p')$, as this would double count this term.  However, in all applications either $k_i$ or $k_f$ is on-shell, so the full kernel contains {\it  no\/} term of the form  $\Theta(k_f)\otimes \Theta(k_i)$, so the expansion
\begin{widetext}
\bea
\widetilde{V}_{\beta\beta',\alpha\alpha'}(k_f,k_i;P)&=&V^{A}_{\beta\beta',\alpha\alpha'}(k_f,k_i;P)
+\Big[{V}^{1i}_{\beta\beta_1,\alpha\alpha'}(k_f,k_i;P)\,\Theta_{\beta_1\beta'}(k_i)
+\Theta_{\beta\beta_1}(k_f)\,{V}^{1f}_{\beta_1\beta',\alpha\alpha'}(k_f,k_i;P) \Big],\qquad \label{eq:Vinreducedform}
\eea
\end{widetext}
is useful, and will be used below.   

It is important to realize that even thought this interaction current was fixed here using the principle of picture independence in the case when $h=1$, it still satisfies the two-body WT identity and hence is an acceptable choice for the interaction current, even in the general case when $h\ne1$.   

Matrix elements of the interaction currents involve integrals over the initial and final three momenta.   Inserting the general result (\ref{eq:Icurrent}) for the interaction current into the interaction current term in Eq.~(\ref{eq:23}), and writing the reduced kernels in terms of their three independent four-vector arguments,  gives the somewhat  simplified result
\begin{widetext}
\bea
\left<{V}^\mu\right>=\int_k\int_{k'} \overline{\it \Psi}_{\lambda_n\alpha}^{\lambda}&&(k,P_+)\,h(p_+)\widetilde{V}^\mu_{\lambda_n\lambda_n',\alpha\alpha'}(k\,P_+; k'\,P_-)\,h(p'_-)\,{\it \Psi}_{\alpha'\lambda_n'}^{\lambda'}(k',P_-) 
\nonumber\\
=\frac1{2m} \int_k\int_{k'}  \overline{\it \Psi}_{\lambda_n\alpha}^{\lambda}&&(k,P_+)\,h(p_+)\Big[{V}^{2i}_{\lambda_n\lambda_n',\alpha\alpha_1}(k,k';P_+)\,j_{0\,\alpha_1\alpha'}^\mu(q) +j_{0\,\alpha\alpha_1}^\mu(q)\, {V}^{2f}_{\lambda_n\lambda_n',\alpha_1\alpha'}(k,k';P_-)
\nonumber\\
+&&{V}^{1i}_{\lambda_n\beta,\alpha\alpha'}(k,k'+q;P_+)\,j_{0\,\beta\lambda'_n}^\mu(q) +j_{0\,\lambda_n\beta}^\mu(q)\, {V}^{1f}_{\beta\lambda_n',\alpha\alpha'}(k-q,k';P_-)\Big]h(p'_-)\,{\it \Psi}_{\alpha'\lambda_n'}^{\lambda'}(k',P_-)\, . \qquad\quad \label{eq:intform2}
\eea
Further insight and a check on these results can be found in Appendix \ref{sec:qto0}, where it is shown how this expression for the exchange current reduces to (\ref{eq:211}) when $q\to0$. 

At this point it is convenient to rewrite the interaction current (\ref{eq:intform2}) in another form which will remove any reference to the truncated kernels, and display the current in a form similar to that found already in diagrams (A) and (B).  
To this end, recall the relativistic wave equations for the bound state, Eq.~(\ref{eq:212a}), and use these  to introduce convenient {\it truncated\/} vertex functions defined by  
%
%
\begin{align}
S^{-1}_{\alpha\alpha'}(p_-){\it \Psi}^{(2) \lambda'}_{\alpha'\lambda_n}(k,P_-)
&=-h(p_-)\,\int_{k'}\Theta_{\alpha\alpha_1}(p_-){V}^{2f}_{\lambda_n\lambda'_n,\alpha_1\alpha'}(k,k'; P_-)\,
h(p'_-)
{\it \Psi}^{\lambda'}_{\alpha'\lambda'_n}(k',P_-) 
\nonumber\\
\overline{\it \Psi}^{(2) \lambda}_{\lambda'_n\alpha'} (k',P_+) S^{-1}_{\alpha'\alpha}(p'_+)
&=-\int_{k} 
\overline{{\it \Psi}}^\lambda_{\lambda_n\alpha}(k,P_+)\,h(p_+)\,
{V}^{2i}_{\lambda_n\lambda'_n,\alpha\alpha''}(k,k'; P_+)\, \Theta_{\alpha''\alpha}(p'_+)\,h(p'_+)
\, .\label{eq:thetawf1}
\end{align}
%
where $\Theta$ restores the factor originally removed from the definition (\ref{eq:reducedk}), so that $\Theta(p_-) {V}^{2f}(k,k';P_-)$ is that part of the kernel proportional to $\Theta(p_-)$. Because $2mS(p_-)=\Theta^{-1}(p_-)$, the factors of $S^{-1}$ and $2m\Theta$ can be dropped from both sides of the equation, giving
%
\begin{align}
2m{\it \Psi}^{(2) \lambda'}_{\alpha\lambda_n}(k,P_-)
&=-h(p_-)\,\int_{k'}{V}^{2f}_{\lambda_n\lambda'_n,\alpha\alpha'}(k,k'; P_-)\,
h(p'_-)
{\it \Psi}^{\lambda'}_{\alpha'\lambda'_n}(k',P_-) 
\nonumber\\
2m \overline{\it \Psi}^{(2) \lambda}_{\lambda'_n\alpha'} (k',P_+) 
&=-\int_{k} 
\overline{{\it \Psi}}^\lambda_{\lambda_n\alpha}(k,P_+)\,h(p_+)\,
{V}^{2i}_{\lambda_n\lambda'_n,\alpha\alpha'}(k,k'; P_+)\,h(p'_+)
\, .\label{eq:thetawf}
\end{align}
%
While these equations are
are simpler to use analytically, the original versions (\ref{eq:thetawf1}) are easier to work with numerically because their kernels are easily recognizable  parts of the full kernel used in the original bound state equations.  The wave function ${\it \Psi}^{(2)}$ 
is a convenient object because it can be computed at the same time the wave function ${\it \Psi}$ is computed, removing all of the details of the OBE model from the computations of the form factors.

Similar equations hold for the  ${V}^{1}$ kernels, but here both nucleons are off-shell, so the equations take a slightly different form, similar to Eqs.~(\ref{eq:28a}) 
\begin{align}
&{\cal G}^{(1) \lambda'}_{\alpha\beta}(k-q,P_-)
=-h(k-q)h(p_+)\,\Theta_{\beta\beta'}(k-q)\int_{k'} {V}^{1f}_{\beta'\lambda_n',\alpha\alpha'}( k-q,   k'; P_-)\,h(p'_-)\,{\it \Psi}^{\lambda'}_{\alpha'\lambda_n'}(k',P_-) \qquad
\nonumber\\
&\overline{\cal G}^{(1) \lambda}_{\beta\alpha'}(k'+q,P_+)
=-\int_{k} \,\overline{\it\Psi}^\lambda_{\lambda_n\alpha}(k',P_+)\,h(p_+)\,{V}^{1i}_{\lambda_n\beta',\alpha\alpha'}( k,k'+q; P_+) \,\Theta_{\beta'\beta}(k'+q)\,h(k'+q)h(p'_-)\, ,
\label{eq:28aaa}
\end{align}
where here it is convenient to work directly with ${\cal G}$.   These equations are rewritten by moving the $\Theta$ projection operators to the other side, and dividing by the strong form factors (converting ${\cal G}\to \widetilde{\cal G}$), giving
\begin{align}
&2m\,S_{\beta\beta'}(k-q) \widetilde{\cal G}^{(1) \lambda'}_{\alpha\beta'}(k-q,P_-) 
=-\int_{k'} {V}^{1f}_{\beta\lambda_n',\alpha\alpha'}( k-q,   k'; P_-)\,h(p'_-)\,{\it \Psi}^{\lambda'}_{\alpha'\lambda_n'}(k',P_-) \qquad
\nonumber\\
&2m\, \overline{\widetilde{\cal G}}^{(1) \lambda}_{\beta'\alpha'}(k'+q,P_+) S_{\beta'\beta}(k+q)
=-\int_{k} \,\overline{\it\Psi}^\lambda_{\lambda_n\alpha}(k',P_+)\,h(p_+)\,{V}^{1i}_{\lambda_n\beta,\alpha\alpha'}( k,k'+q; P_+)\, .
\label{eq:28ab}
\end{align}
As shown below, the vertex function ${\cal G}^{(1)}$ will soon be replaced by another more useful object. 

Substituting (\ref{eq:thetawf}) and (\ref{eq:28ab}) into (\ref{eq:intform2}), the matrix element of the interaction current breaks into two terms:
%
\bea
\left<V^\mu\right>&=&\left<V^\mu_2\right>+\left<V^\mu_1\right>
\eea
where
\bea
\left<V^\mu_2\right>&=&- \int_k\Bigg\{ \overline{\it \Psi}_{\lambda_n\alpha}^{\lambda}(k,P_+)\,\frac{h_+}{h_-} j^\mu_{0\, \alpha\alpha'}(q){\it \Psi}_{\alpha'\lambda_n}^{(2) \lambda'}(k,P_-)
+\overline{\it \Psi}_{\lambda_n\alpha}^{(2) \lambda}(k,P_+)\,\frac{h_-}{h_+} j^\mu_{0\,\alpha\alpha'}(q)\,{\it \Psi}_{\alpha'\lambda_n}^{\lambda'}(k,P_-) 
\Bigg\}
\nonumber\\
\left<V^\mu_1\right>&=&- \int_k\overline{\widetilde{\cal G}}_{\lambda_n\alpha}^{\lambda}(k,P_+)\Big[h^2_+S_+\Big]_{\alpha\alpha'} \widetilde{\cal G}^{(1) \lambda'}_{\alpha' \beta'}(k-q,P_-)\,j_{0\,\lambda_n\beta}^\mu(q)\,S_{\beta\beta'}(k-q)
\nonumber\\
&&\qquad\qquad+\overline{\widetilde{\cal G}}_{\beta'\alpha'}^{(1) \lambda}(k+q,P_+)\Big[h^2_-S_-\Big]_{\alpha\alpha'}\widetilde{\cal G}^{\lambda'}_{\alpha' \lambda'_n}(k,P_-)\,S_{\beta'\beta}(k+q)\,j_{0\,\beta\lambda'_n}^\mu(q)\Big]\, , \qquad\quad
\eea
with $h_\pm=h(p_\pm)$ and $S_\pm=S(p_\pm)$, and the order of some of the terms has been interchanged, possible because the indices are shown on all matrices.  
The matrix element $\left<V^\mu_2\right>$ is expressed in terms of wave functions, ${\it \Psi}=S\,{\cal G}$, while the matrix element $\left<V^\mu_1\right>$ is expressed entirely in terms of reduced vertex functions, $\widetilde{\cal G}$.  This facilitates comparison with the expressions for diagrams (A) and (B).  
Replacing the sum over on-shell spinors by the projection operator, and removing the charge conjugation matrices permits each of these matrix elements to be written as a trace.  The result is
\begin{subequations}
\bea
\left<V^\mu_2\right>&=&\int_k{\rm tr}\Bigg[ \Big\{
\overline{\Psi}^{\lambda}(k,P_+)\,\frac{h_+}{h_-} j_{0}^\mu(q) {\Psi}^{(2) \lambda'}(k,P_-)+
\overline{\Psi}^{(2) \lambda}(k,P_+)\,j_{0}^\mu(q)\frac{h_-}{h_+}\,{\Psi}^{\lambda'}(k,P_-)\Big\}\,\Lambda(-k) \Bigg] \label{eq:V2}
\\
\left<V^\mu_1\right>&=&\int_k{\rm tr}\,\Big[\overline{\widetilde\Gamma}^{\lambda}(k,P_+)h^2_+S_+\widetilde{\Gamma}^{(1) \lambda'}(k-q,P_-)S(-k+q)\,j_{0}^\mu(q)\,\Lambda(-k) 
\nonumber\\&&\qquad\qquad
+\overline{\widetilde\Gamma}^{(1) \lambda}(k+q,P_+)h^2_-S_-\widetilde{\Gamma}^{\lambda'}(k,P_-)\,\Lambda(-k)\,j_{0}^\mu(q)\,S(-k-q)\Big] \label{eq:V1a}
\\
&=& \int_k\frac{mE_k}{{\bf k}\cdot{\bf q}}{\rm tr}\,\Bigg\{\frac{1}{k_0}\overline{\widetilde\Gamma}^{\lambda}(\tilde k_+,P_+)\,S_d(\tilde p)\,\widetilde{\Gamma}^{(1) \lambda'}(\tilde k_-,P_-)\,\Lambda(-\tilde k_-)\,j_{0}^\mu(q)\,\Lambda(-\tilde k_+)\Big|_{k_0=E_+}
\nonumber\\&&\qquad\qquad\qquad
-\frac{1}{k_0}\overline{\widetilde\Gamma}^{(1) \lambda}(\tilde k_+,P_+)\,S_d(\tilde p)\,\widetilde{\Gamma}^{\lambda'}(\tilde k_-,P_-)\,\Lambda(-\tilde k_-)\,j_{0}^\mu(q)\,\Lambda(-\tilde k_+)\Big|_{k_0=E_-}\Bigg\} 
\, , \label{eq:V1}
\eea
\end{subequations}
where the result (\ref{eq:V1}) for $\left<V^\mu_1\right>$ was obtained by shifting ${\bf k}\to {\bf k}+\frac12{\bf q}$ in the first term and ${\bf k}\to {\bf k}-\frac12{\bf q}$ in the second, recalling that $S_d(\tilde p)=h^2(\tilde p)S(\tilde p)$, and using (\ref{eq:Skpkm}) and the notation of Eq.~(\ref{eq:ktilde}).  While the expression (\ref{eq:V1}) for $\left<V^\mu_1\right>$ looks similar to the expression (\ref{eq:Bpmterms3}) for $J^\mu_{{\rm B}_\pm}$, it differs in the  fact that the two factors in the brackets $\{\,\}$ do not cancel when $q\to0$ as they do in (\ref{eq:Bpmterms3}).  In (\ref{eq:V1})  the  apparent singularity at $q\to0$ is cancelled by the behavior of the terms $\widetilde\Gamma^{(1)}(k,P)\Lambda(-k)$, each of which vanish when $q\to0$. 

\subsection{Combined result} \label{sec:combined}

The results for the interaction currents can now be combined with the formulae for the (A) and (B) diagrams, displaying the cancellations between the interaction current and contributions from the (B) diagrams, discussed diagramatically in Sec.~\ref{sec:rules}.  Combining (\ref{eq:Atrace}) and (\ref{eq:V2}) gives
\bea
J^\mu_{\lambda\lambda'}(q)\Big|_{{\rm A}+V_2}=-\int_k {\rm tr}\Bigg[\Big\{&&\overline{\Psi}^\lambda (k,P_+)j^\mu(p_+,p_-)\Psi^{\lambda'}(k,P_-)
\nonumber\\&&
-\overline{\Psi}^\lambda (k,P_+)\frac{h_+}{h_-} j_{0}^\mu(q) {\Psi}^{(2) \lambda'}(k,P_-)
-\overline{\Psi}^{(2) \lambda}(k,P_+)\,j_{0}^\mu(q)\frac{h_-}{h_+}\Psi^{\lambda'}(k,P_-)
\Big\}\Lambda(-k)\Bigg]\, . \qquad\label{eq:A&V2}
\eea
Combining (\ref{eq:Bpmterms3}) and (\ref{eq:V1}) gives
\bea
J^\mu_{\lambda\lambda'}(q)\Big|_{{\rm B}+V_1}=\int_k&&\left[\frac{m E_k}{{\bf k}\cdot{\bf q}}\right]{\rm tr} \Bigg\{\frac1{k_0}\,
\overline{\widehat\Gamma}^{\lambda}_{BS}(\widetilde k_+,P_+)\, S_d(\widetilde p)\,\widetilde\Gamma^{\lambda'}(\widetilde k_-,P_-)\Lambda(-\widetilde k_-) \,j_0^\mu(q)\,\Lambda(-\widetilde k_+) \Big|_{k_0=E_-}
\nonumber\\&& 
-\frac1{k_0}\, \overline{\widetilde\Gamma}^\lambda(\widetilde k_+,P_+)\, S_d(\widetilde p)\,\widehat\Gamma^{\lambda'}_{BS}(\widetilde k_-,P_-)\Lambda(-\widetilde k_-) \,j^\mu_0(q)\,\Lambda(-\widetilde k_+)\Big|_{k_0=E_+} \Bigg\}\, .\quad
\label{eq:B&V1}
\eea
\end{widetext}
where we introduced a new vertex function
\bea
\widehat{\Gamma}^\lambda_{BS}(\widetilde k,P)\equiv \widetilde\Gamma^{\lambda}(\widetilde k,P)-\widetilde\Gamma^{(1) \lambda}(\widetilde k,P)
\eea
which is the result of the cancellations between the (B) diagram contributions, $\widetilde \Gamma$, and the exchange currents arising from particle 1, $\widetilde \Gamma^{(1)}$.   Using the subscript ``BS''  on this amplitude reminds us that it is a vertex functions with {\it both\/} of the final nucleons are off-shell, and hence has the structure of a Bethe-Salpeter  vertex function.  However, as discussed in Sec.\ \ref{sec:BScancellation}, this is obtained by quadratures from the CST wave functions, and hence would differ from that obtained from the solution of a BS equation.  

The complete result for the deuteron current is the sum of the two terms  (\ref{eq:A&V2}) and (\ref{eq:B&V1}).
This is the final result of this paper; further reductions of these results must await the discussions in Ref.\ II. 

\subsection{Computation of the BS vertex functions}  \label{sec:BScancellation}

 The BS vertex functions, $\widehat \Gamma_{BS}$, that enter into 
 Eq.\ (\ref{eq:B&V1}) are computed from a generalization of Eq.~(\ref{eq:28a}).  Going to the rest frame, removing the on-shell spinors, and writing the result for the reduced vertex function allows (\ref{eq:28a}) to be written in the form
 \begin{widetext}
\bea
&&\widetilde{\cal G}^{\lambda'}_{\alpha\beta}(k_f,P)=\Big[\widetilde{\Gamma}_{BS}^{\lambda'}(k_f,P){\cal C}\Big]_{\alpha\beta}
=-\int_{k'} \widetilde{V}_{\beta\beta'',\alpha\alpha'}( k_f,  k'; P)\,h(p')
{\it\Psi}^{\lambda'}_{\alpha'\beta'}(k',P) \Lambda^T_{\beta'\beta''}(k')\qquad\quad
\nonumber\\
&&\overline{\widetilde{\cal G}}^{\lambda}_{\beta'\alpha'}(k_i',P)=\Big[{\cal C}\,\overline{\widetilde{\Gamma}}_{BS}^{\lambda}(k_i',P)\Big]_{\beta'\alpha'}
=-\int_{k}\Lambda^T_{\beta''\beta}(k)\overline{{\it\Psi}}^{\lambda}_{\beta\alpha}(k,P)  h(p) \widetilde{V}_{\beta''\beta',\alpha\alpha'}( k,  k_i'; P)\,\qquad\quad
\label{eq:offshellG}
\eea
where, reflecting the fact that these equations are used under a double integral, the role of $k$ and $k'$ and the indices have been interchanged in the two equations, and the off-shell momenta are distinguished by another subscript: $k_f$ or $k'_i$. The off-shell dependence in these equations  comes entirely from the kernel; the wave function under the integral (which depends on $k'^2=m^2$ or $k^2=m^2$) contains none of this dependence.  Multiplying (\ref{eq:28ab}) by $\Theta$ gives  the the $\widehat \Gamma$'s needed in (\ref{eq:B&V1})  
\begin{subequations}
\bea
\Big[\widehat\Gamma^{\lambda'}_{BS}(k_f,P){\cal C}\Big]_{\alpha\beta}&\equiv&\Big[\big(\widetilde{\Gamma}^{\lambda'}(k_f,P)-{\Gamma}^{(1)\lambda'}(k_f,P)\big){\cal C}\Big]_{\alpha\beta}
\nonumber\\
&=&-\int_{k'}\Big[ \widetilde{V}_{\beta\beta'',\alpha\alpha'}( k_f,   k'; P)-\Theta_{\beta\beta_1}(k_f){V}^{1f}_{\beta_1\beta'',\alpha\alpha'}(k_f,k';P)\Big]\,h(p')
{\it \Psi}^{\lambda'}_{\alpha'\beta'}(k',P)\Lambda^T_{\beta'\beta''}(k') \qquad\quad
\nonumber\\
&=&-\int_{k'}\widehat{V}_{\beta\beta'',\alpha\alpha'}( k_f,   k'; P)\,h(p')
{\it\Psi}^{\lambda'}_{\alpha'\beta'}(k',P)\Lambda^T_{\beta'\beta''}(k') \label{eq:BScancellationa}
\\
\Big[{\cal C}\overline{\widehat\Gamma}^{\lambda}_{BS}(k_i',P)\Big]_{\beta'\alpha'}&\equiv&\Big[{\cal C}\big(\overline{\widetilde{\Gamma}}^{\lambda}(k_i',P)-\overline{{\Gamma}}^{(1)\lambda}(k_i',P)\big)\Big]_{\beta'\alpha'}
\nonumber\\
&=&-\int_{k}\Lambda^T_{\beta''\beta}(k){\it \overline{\Psi}}^{\lambda}_{\beta\alpha}(k,P)\,h(p)\Big[ \widetilde{V}_{\beta''\beta',\alpha\alpha'}( k,   k_i'; P)-{V}^{1i}_{\beta''\beta_1,\alpha\alpha'}(k,k_i';P)\Theta_{\beta_1\beta'}(k_i')\Big]
 \qquad\quad
\nonumber\\
&=&-\int_{k}\Lambda^T_{\beta''\beta}(k){\it \overline{\Psi}}^{\lambda}_{\beta\alpha}(k,P)\,h(p) \widehat{V}_{\beta''\beta',\alpha\alpha'}( k,   k_i'; P)\label{eq:BScancellation}
\eea
\end{subequations}
\end{widetext}
where $\widehat V$ is the kernel {\it without\/} any off-shell projection operators $\Theta(k)$ in the numerator (or $\Theta(k')$ for the conjugate equation), and $\widehat\Gamma_{BS}$ is the vertex function computed from this kernel.  Since there is no $\Theta$  associated with the on-shell particle, and the $\Theta$  factor that would  be present when the particle is off-shell has been cancelled, the expansions (\ref{eq:Vinreducedform}) show immediately that [returning to the notation of Eq.~(\ref{eq:Vinreducedform}) ]
\bea
\widehat V_{\beta\beta',\alpha\alpha'}(k_f,k_i;P)=V^{A}_{\beta\beta',\alpha\alpha'}(k_f,k_i;P)\, , \label{eq:hatkernels}
\eea
where $V^{A}$ is defined by the expansion (\ref{eq:Vinreducedform}), and either $k_f$ or $k_i$ may be off-shell, but not both.
The $\widehat\Gamma_{BS}$ vertex functions have the full BS structure (depending on two four-momenta that are both off-shell), but, as previously emphasized, this dependence arises  only from the dependence of the truncated kernels (\ref{eq:hatkernels}) on the off-shell four-momenta $k_f$ (or $k_i$) and not on the wave functions ${\it \Psi}$ (or ${\it \bar \Psi}$) from which they are calculated.  

\acknowledgements
It is a pleasure to acknowledge helpful conversations with J.~W.~Van Orden, who also contributed to the early stages of this work.  This work was partially support by Jefferson Science
Associates, LLC, under U.S. DOE Contract No. DE-AC05-
06OR23177.


\appendix

\section{Derivation of the two-body WT identity} \label{app:WTidet}

Using the WT identity (\ref{eq:WTdressed}) for the dressed single nucleon current, and recalling that the wave functions and vertex functions include the nucleon form factors, the divergence of the  current (\ref{eq:23}) reduces to
\begin{widetext}
\begin{align}
q_\mu J^\mu(q)=&\int_k \Bigg\{
	 \overline{\it \Psi}^\lambda_{\lambda_n\alpha}(k,P_+)\,e_0\Big[\frac{h(p_+)}{h(p_-)}S^{-1}(p_-)-\frac{h(p_-)}{h(p_+)}S^{-1}(p_+)\Big]_{\alpha\alpha'}\,{\it\Psi}^{\lambda'}_{\alpha'\lambda_n}(k,P_-)
 \nonumber\\
 &+\int_{k'}
 \overline{\it\Psi}^\lambda_{\lambda_n\alpha}(k,P_+)\,\Big[q_\mu V^\mu_{\lambda_n\lambda_n',\alpha\alpha'}(k\,P_+;k'\,P_-)\Big]\,{\it\Psi}^{\lambda'}_{\alpha'\lambda'_n}(k',P_-)\Bigg\}
\nonumber\\
&+e_0\int_{k_+} \overline{\it\Psi}^\lambda_{\lambda_n\alpha}(\hat k_+,P_+)\, \frac{{\cal G}^{\lambda'}_{\alpha\beta}(k_-,P_-)}{h(k_-)}\,\bar{u}^T_\beta({\bf k}_+,\lambda_n)
-e_0\int_{k_-} u^T_{\beta'}({\bf k}_-,\lambda_n)\, \frac{\overline{\cal G}^\lambda_{\beta'\alpha'}(k_+,P_+)}{h(k_+)}{\it\Psi}^{\lambda'}_{\alpha'\lambda_n}(\hat k_-,P_-) \, . \label{eq:25}
\end{align}
%
Note that the last two integrals are over ${\bf k}_\pm$ [with the $E_k\to E_\pm$ in (\ref{eq:volint1})], and the  transpose label is retained on the spinors $\bar u$ and $u$ in the last two terms, as would be required if we were to drop the Dirac indices [with the indices shown, the transpose symbol is not necessary].

The requirement that the current be conserved ($q_\mu J^\mu(q)=0$) converts Eq.~(\ref{eq:25}) into a constraint on the divergence of the interaction current.   In the first term [arising from the divergence of diagram (A)] we use the two-body CST equation (\ref{eq:bsequation}) to remove the inverse propagators.  Writing this equation and its conjugate in terms of the {\it reduced\/} kernel defined in Eq.~(\ref{eq:VandVmu})
\begin{align}
S^{-1}_{\alpha\alpha'}(p_-)\,{\it\Psi}^{\lambda'}_{\alpha'\lambda_n}(k,P_-)
&=-h(p_-)\int_{k'} \widetilde{V}_{\lambda_n\lambda'_n,\alpha\alpha'}(k,k'; P_-)\,h(p'_-)
{\it\Psi}^{\lambda'}_{\alpha'\lambda'_n}(k',P_-)
\nonumber\\
\overline{\it\Psi}^\lambda_{\lambda'_n\alpha}(k',P_+)\,S^{-1}_{\alpha\alpha'}(p'_+)
&=-\int_{k} 
\overline{\it\Psi}^\lambda_{\lambda_n\alpha}(k,P_+)\,h(p_+)\,\widetilde{V}_{\lambda_n\lambda'_n,\alpha\alpha'}(k,k'; P_+)\,h(p'_+)
\, ,\label{eq:212a}
\end{align}
pulling out the on-shell spinors from the kernels from $\widetilde{V}_{\lambda_n\lambda'_N,\alpha\alpha'}$ using
\bea
\widetilde{V}_{\lambda\lambda',\alpha\alpha'}(k,k';P)=\bar u_\beta({\bf k},\lambda)\,\widetilde{V}_{\beta\beta',\alpha\alpha'}(k,k';P)\,u_{\beta'}({\bf k}',\lambda'),
\eea
gives the following contribution from diagram (A)
\bea
q_\mu J^\mu(q)\Big|_{\rm (A)}=e_0\int_k \int_{k'}
 \overline{\it\Psi}^\lambda_{\lambda_n\alpha}(k,P_+)\,\overline{u}_\beta({\bf k},\lambda_n)h(p_+)\,\Delta\widetilde{V}^{\rm (A)}_{\beta\beta',\alpha\alpha'}(kP_+;k' P_-)\,h(p'_-)\,u_{\beta'}({\bf k}',\lambda'_n)
{\it\Psi}^{\lambda'}_{\alpha'\lambda_n'}(k',P_-).
\qquad \label{eq:2.13}
\eea
with
\bea
\Delta\widetilde{V}^{\rm (A)}_{\beta\beta',\alpha\alpha'}(kP_+;k' P_-)\equiv\widetilde{V}_{\beta\beta',\alpha\alpha'}(k,k'; P_+)-\widetilde{V}_{\beta\beta',\alpha\alpha'}(k,k'; P_-).
\eea
Note that, if the reduced kernel is {\it independent of the total four-momentum $P$\/} (not generally the case), diagram (A) is conserved.

In the last two terms of (\ref{eq:25}) [which arise from the (B) diagrams] we shift ${\bf k}_+\to{\bf k}$ in the first term (which also shifts $\hat k_+\to k$ and $k_-\to k-q$), and ${\bf k}_-\to{\bf k}$ in the second term (which also shifts $\hat k_-\to k$ and $k_+\to k+q$), giving
\bea
q_\mu J^\mu(q)\Big|_{{\rm (B)}}=e_0\Bigg\{\int_{k} \overline{\it\Psi}^\lambda_{\lambda_n\alpha}(k,P_+)\, \frac{{\cal G}^{\lambda'}_{\alpha\beta}(k-q,P_-)}{h(k-q)}\,\bar{u}^T_\beta({\bf k},\lambda_n)
- u^T_{\beta'}({\bf k},\lambda_n)\, \frac{\overline{\cal G}^\lambda_{\beta'\alpha'}(k+q,P_+)}{h(k+q)}{\it\Psi}^{\lambda'}_{\alpha'\lambda_n}(k,P_-)\Bigg\}\, . \label{eq:Bpm2}
\eea
Next, use the generalization of (\ref{eq:212a}) to the cases when both nucleons are off-shell to express ${\cal G}$ in terms of $V{\it \Psi}$.  The equations we need, factoring out the on-shell spinors, are
\begin{align}
&{\cal G}^{\lambda'}_{\alpha\beta}(k-q,P_-)\,\bar{u}^T_\beta({\bf k},\lambda_n)
=-h(k-q)h(p_+)\int_{k'} \,\bar{u}_\beta({\bf k},\lambda_n)\widetilde{V}_{\beta\beta',\alpha\alpha'}( k-q,   k'; P_-)\,h(p'_-)\,u_{\beta'}({\bf k}',\lambda_n')
{\it \Psi}^{\lambda'}_{\alpha'\lambda_n'}(k',P_-) \qquad
\nonumber\\
&{u}^T_{\beta'}({\bf k},\lambda')\,\overline{\cal G}^\lambda_{\beta'\alpha'}(k'+q,P_+)
=-\int_{k} \,\overline{\it\Psi}^\lambda_{\alpha\lambda}(k',P_+)\bar u_\beta({\bf k},\lambda_n)\,h(p_+)\,\widetilde{V}_{\beta\beta',\alpha\alpha'}( k,k'+q; P_+){u}_{\beta'}({\bf k},\lambda_n') \,h(k'+q)h(p'_-)
\label{eq:28a}
\end{align}
where $p_\pm=P_\mp-(k \mp q)=P_\pm-k$ for both primed and unprimed variables.  
In both equations the transpose label on the spinors on the left-hand side of each equation can be dropped because the order of the terms is already appropriate for matrix multiplication.  Substituting the relations (\ref{eq:28a}) into (\ref{eq:Bpm2})  makes it possible to write the divergence of the interaction current as an operator relation.  Requiring $q_\mu J^\mu(q)=0$, and reorganizing Eq.~(\ref{eq:25}), gives 
\begin{align}
 \int_k\int_{k'} 
 \overline{\it\Psi}^\lambda_{\lambda\alpha}(k,P_+)\,\bar u_\beta({\bf k},\lambda)&\,h(p_+)\Big[q_\mu \widetilde{V}^\mu_{\beta\beta',\alpha\alpha'}(k\,P_+;k'\,P_-)\Big]h(p'_-)\,u_{\beta'}({\bf k}',\lambda')\,{\it\Psi}^{\lambda'}_{\alpha'\lambda'}(k',P_-)
\nonumber\\
=e_0 \int_k\int_{k'} 
 \overline{\it\Psi}^{\lambda}_{\lambda\alpha}(k,P_+)\,&\bar u_\beta({\bf k},\lambda)
 \,h(p_+)
  \nonumber\\
&\times\bigg[\Delta \widetilde{V}^{\rm (A)}_{\beta\beta',\alpha\alpha'}(k P_+;k' P_-)  -\Delta \widetilde{V}^{\rm (B)}_{\beta\beta',\alpha\alpha'}(k P_+;k' P_-)\bigg]h(p'_-)\, u_\beta({\bf k}',\lambda'){\it\Psi}^{\lambda'}_{\alpha'\lambda'}(k',P_-) \, , \label{eq:28}
\end{align}
where the new $\Delta \widetilde{V}$ from diagram (B) is
\bea
\Delta \widetilde{V}^{\rm (B)}_{\beta\beta',\alpha\alpha'}(k P_+;k' P_-)\equiv  
\widetilde{V}_{\beta\beta',\alpha\alpha'}(k_,k'+q;P_+)
-\widetilde{V}_{\beta\beta',\alpha\alpha'}(k-q,k';P_-)
\eea
and in every term $k^2=m^2$ and $k'^2=m^2$, so that the off-shell momenta are clearly identified.
In operator form this gives Eq.~(\ref{eq:29}), one form of the two-body WT identity.

Alternatively,  writing the kernels in terms of the four-momenta of particles 1 and 2, $\{k,p\}$, in both the initial and final state, so that the {\it three independent\/} momenta are expressed in terms of {\it four dependent\/} momenta
\bea
V(k,k';P)\equiv V(k,p;k',p')
\eea
the identity  (\ref{eq:29}) can be written in a second form

\begin{figure*}
\centerline{
\mbox{
\includegraphics[width=6in]{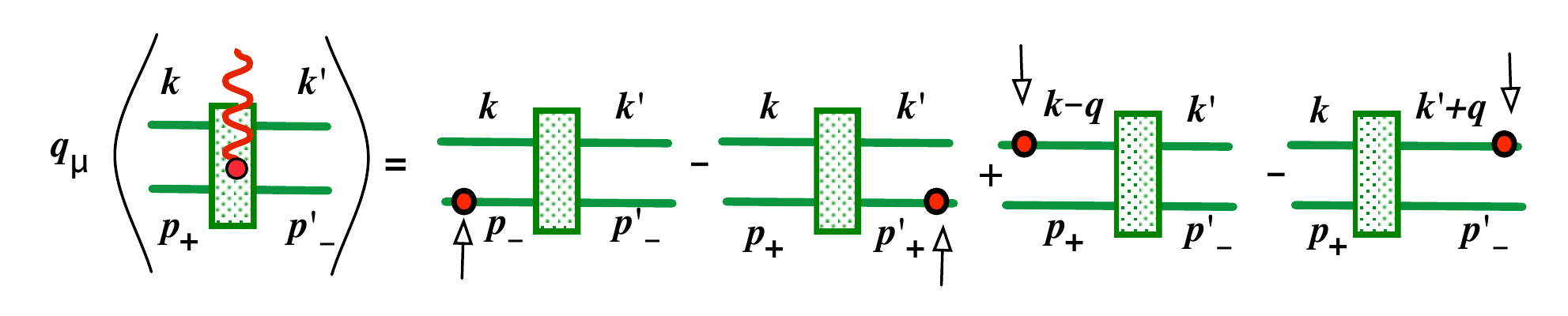}
}
}
\caption{\footnotesize\baselineskip=10pt (Color on line) Diagramatic representation of the two body WT identity (\ref{eq:210}).  Momentum $q$ flows into each diagram (as illustrated by the arrow) at the location of the red dot.  The momenta are defined in the text.  }
\label{Fig2}
\end{figure*}

%
\begin{align}
q_\mu \widetilde{V}^\mu_{\beta\beta',\alpha\alpha'}(k,p_+;k',p'_-)
=&e_0\Big[\widetilde{V}_{\beta\beta',\alpha\alpha'}(k, p_-;k', p'_-)-\widetilde{V}_{\beta\beta',\alpha\alpha'}(k\,p_+;k', p'_+)
\nonumber\\& 
+ \widetilde{V}_{\beta\beta',\alpha\alpha'}(k-q, p_+;k', p'_-)
 -\widetilde{V}_{\beta\beta',\alpha\alpha'}(k, p_+;k'+q, p'_-)\Big] \label{eq:210}
\end{align}
which is illustrated in Fig.~\ref{Fig2}.


\section{Evaluation of the derivative term in Eq.\ (\ref{eq:212})} \label{app:A}

Expanding the derivative term in Eq.\ (\ref{eq:212}) (relabeling $\widetilde k\to k$ and $\widetilde p\to p$ for convenience and recalling that $\partial/\partial k_0=-\partial/\partial p_0$)

\begin{align}
J^0_{\lambda\lambda'}(0)\Big|_{{\rm B}_\pm}=&  \int_k \frac{m}{E_k}  \Bigg\{\overline{\widetilde{{\Gamma}}}^{\lambda}_{\beta\alpha}(k,P)\,\frac{\partial}{\partial p_0}\big[h^2(p)S(p)\big]_{\alpha\alpha'}\widetilde{\Gamma}^{\lambda'}_{\alpha'\beta'}(k,P)\,\Big[\Lambda(-k)\,e_0\gamma^0\,\Lambda(-k)\Big]_{\beta\beta'}
\nonumber\\
&- \frac{\partial}{\partial k_0}\left[\overline{\widetilde{\Gamma}}^{\lambda}_{\beta\alpha}(k,P)\right] \, \big[h^2(p)S(p)\big]_{\alpha\alpha'}\,\widetilde{\Gamma}^{\lambda'}_{\alpha'\beta'}(k,P)\,\Big[\Lambda(-k)\,e_0\gamma^0\,\Lambda(-k)\Big]_{\beta\beta'}
\nonumber\\
&- \overline{\widetilde{\Gamma}}^{\lambda}_{\beta\alpha}(k,P)\,\big[h^2(p)S(p)\big]_{\alpha\alpha'}\,\frac{\partial}{\partial k_0}\Big[\widetilde{\Gamma}^{\lambda'}_{\alpha'\beta'}(k,P)\Big]\,\Big[\Lambda(-k)\,e_0\gamma^0\,\Lambda(-k)\Big]_{\beta\beta'} 
\nonumber\\
&-E_k \,\overline{\widetilde{\Gamma}}^{\lambda}_{\beta\alpha}(k,P)
\,\big[h^2(p) S(p)\big]_{\alpha\alpha'}\,\widetilde{\Gamma}^{\lambda'}_{\alpha'\beta'}(k,P) \,\frac{\partial}{\partial k_0} \bigg[\frac{
\Lambda(-k)\,e_0\gamma^0\,\Lambda(-k)}{k_0}\bigg]_{\beta\beta'}
\Bigg\} \Bigg|_{k^2=m^2}\, .
\label{eq:217}
\end{align}
The first term gives a contribution equal to the RIA.  To see this, recall Eq.~(\ref{eq:33}) and note that
\bea
\frac{\partial}{\partial p_0}\big[h^2(p)S(p)\big]&=&\frac{\partial}{\partial p_0}S_d(p)=S_d(p)\frac{1}{h^2}\Big[\gamma^0 +\frac{2m}{h^2}\frac{\partial h^2}{\partial p_0}\Theta(p) \Big]S_d(p)=\left(\frac1{e_0}\right)h^2(p)\,S(p)j^0(p,p)S(p) \label{eq:dprop}
\, ,
\eea
and use (true when $k^2=m^2$)
\bea
\Big[\Lambda(-k)\,\gamma^0\,\Lambda(-k)\Big]_{\beta\beta'}=-\frac{E_k}{m}\, \Lambda_{\beta\beta'}(-k)\, ,
\,\label{eq:sjs}
\eea
to reduce the first term to (absorbing one factor of $h(p)$ into $\widetilde{\Gamma}$ converting it to ${\Gamma}$) 
\bea
J^0_{\lambda\lambda'}(0)\Big|_{{\rm B}_\pm}^{1^{\rm st} {\rm term}}=-\int_k \overline{\Gamma}^{\lambda}_{\beta\alpha}(k,P)\,\big[S(p)j^0(p,p)S(p)\big]_{\alpha\alpha'}\,{\Gamma}^{\lambda'}_{\alpha'\beta'}(k,P)\Lambda_{\beta'\beta}(-k)
\eea
which is identical to the RIA [Eq.~(\ref{eq:Atrace}) when $q\to0$], if we recall the definition (\ref{eq:27a}).  

Next show that the last term is zero.  For $k^2\ne m^2$, use
\bea
\Lambda(-k)\gamma^0\Lambda(-k)&=&-\frac{k_0}{m}\Lambda(-k)+\frac{m^2-k^2}{4m^2}\gamma^0\qquad \label{eq:219}
\eea
so that
\bea
\frac{\partial}{\partial k_0}\Big[\frac{\Lambda(-k)\,\gamma^0\,\Lambda(-k)}{k_0}\Big]\Big|_{k^2=m^2}=\frac{\gamma^0}{2m^2}-\frac{\gamma^0}{2m^2}=0\, .
\eea

Finally, using (\ref{eq:sjs}) the remaining two terms reduce to 
\begin{align}
J^0_{\lambda\lambda'}(0)\Big|_{{\rm B}_\pm}^{2+3\, {\rm terms}}= e_0  \int_k  \Bigg\{&\frac{\partial}{\partial k_0}\left[\overline{\widetilde{\Gamma}}^{\lambda}_{\beta\alpha}(k,P)\right]
\,h^2(p)S_{\alpha\alpha'}(p)\,\widetilde{\Gamma}^{\lambda'}_{\alpha'\beta'}(k,P)\,\Lambda_{\beta'\beta}(-k)
\nonumber\\
&+ \overline{\widetilde{\Gamma}}^{\lambda}_{\beta\alpha}(k,P)\,h^2(p)S_{\alpha\alpha'}(p)\,
\frac{\partial}{\partial k_0}\Big[\widetilde{\Gamma}^{\lambda'}_{\alpha'\beta'}(k,P)\Big]\,\Lambda_{\beta'\beta}(-k)
 \Bigg\} \Bigg|_{k^2=m^2}\, .
\label{eq:217a}
\end{align}
Now use (\ref{eq:spindecom}) and the properties of the charge conjugation matrix to replace the projection operator $\Lambda(-k)$ by the sum over positive energy spinors
\bea
\Lambda_{\beta'\beta}(-k)=-{\cal C}_{\beta'\gamma'}\Lambda^T_{\gamma'\gamma}(k)\,\overline{\cal C}_{\gamma\beta}
=-\sum_{\lambda_n}{\cal C}_{\beta'\gamma'}u^T_{\gamma'}({\bf k},\lambda_n)\bar u^T_\gamma({\bf k},\lambda_n) \, {\cal C}_{\gamma\beta}
\eea 
This expression can be written
\begin{align}
J^0_{\lambda\lambda'}(0)\Big|_{{\rm B}_\pm}^{2+3\, {\rm terms}}=- e_0  \int_k  \Bigg\{&\frac{\partial}{\partial k_0}\left[\overline{\widetilde{\cal G}}^{\lambda}_{\lambda_n\alpha}(k,P)\right]
\,h^2(p)S_{\alpha\alpha'}(p)\,\widetilde{\cal G}^{\lambda'}_{\alpha'\lambda_n}(k,P)
\nonumber\\&
+ \overline{\widetilde{\cal G}}^{\lambda}_{\lambda_n\alpha}(k,P)\,h^2(p)S_{\alpha\alpha'}(p)\,
\frac{\partial}{\partial k_0}\Big[\widetilde{\cal G}^{\lambda'}_{\alpha'\lambda_n}(k,P)\Big]
 \Bigg\} \Bigg|_{k^2=m^2}\, .
\label{eq:217ab}
\end{align}
Now, use Eq.~(\ref{eq:28a}) for the vertex functions (removing external factors of $h$ and relabeling some of the variables with an eye to the final result)
\begin{align}
&\widetilde{\cal G}^{\lambda'}_{\alpha\lambda_n}(k,P)
=-\int_{k'} \widetilde V_{\lambda_n\lambda_n',\alpha\alpha'}(k,k'; P)\,
h(p')\,{\it \Psi}^{\lambda'}_{\alpha'\lambda_n'}(k',P) 
\nonumber\\
&\overline{\widetilde{\cal G}}^{\lambda}_{\lambda_n'\alpha'}(k',P)
=-\int_{k} \overline{{\it \Psi}}^{\lambda}_{\lambda_n\beta}(k,P)\,h(p)\,\widetilde V_{\lambda_n\lambda_n',\alpha\alpha'}(k,k'; P)\, ,
\label{eq:xx}
\end{align}
 to rewrite (\ref{eq:217ab})
\begin{align}
J^0_{\lambda\lambda'}(0)\Big|_{{\rm B}_\pm}^{2+3\, {\rm terms}}= e_0\int_k\int_{k'} \overline{{\it \Psi}}^\lambda_{\lambda_n\alpha} (k,P)\, h(p)\bigg[\bigg(\frac{\partial }{\partial k'_0}+\frac{\partial}{\partial k_0}\bigg)\widetilde V_{\lambda_n\lambda_n',\alpha\alpha'}(k,k';P)
\bigg]_{\alpha\alpha'}h(p')\,{\it \Psi}^{\lambda'}_{\alpha'\lambda'}(k',P)  \, .
\label{eq:312}
\end{align}
%
Note that these terms cancel the corresponding derivatives in the interaction current giving the result reported in Eq.\ (\ref{eq:212}).


\section{Alternative treatments of the normalization condition}  \label{app:norm}

Expanding the first term in Eq.~(\ref{eq:315}) using  (\ref{eq:33a})  gives
\bea
2e_0 \int_k &&\overline{\it\Psi}^{\lambda}_{\lambda_n\alpha} (k,P)\,\Big\{\gamma^0_{\alpha\alpha'}+ \frac{4m}{h}\frac{\partial h}{\partial P_0}\Theta_{\alpha\alpha'}(p)\Big\}\,{\it\Psi}^{\lambda'}_{\alpha'\lambda_n}(k,P)
\nonumber\\
=&&2e_0 \int_k\Big\{ \overline{\it\Psi}^{\lambda}_{\lambda_n\alpha} (k,P)\,\gamma^0_{\alpha\alpha'}\,{\it\Psi}^{\lambda'}_{\alpha'\lambda_n}(k,P) +\frac{1}{h}\frac{\partial h}{\partial P_0}\Big( \overline{\it\Psi}^{\lambda}_{\lambda_n\alpha} (k,P){\cal G}^{\lambda'}_{\alpha\lambda_n}(k,P)+ \overline{\cal G}^{\lambda}_{\lambda_n\alpha} (k,P){\it \Psi}^{\lambda'}_{\alpha\lambda_n}(k,P)\Big)\Big\}
\nonumber\\
=&&2e_0 \int_k\Big\{ \overline{\it\Psi}^{\lambda}_{\lambda_n\alpha} (k,P)\,\gamma^0_{\alpha\alpha'}\,{\it\Psi}^{\lambda'}_{\alpha'\lambda_n}(k,P) 
-\int_{k'}\Big( \overline{\it\Psi}^{\lambda}_{\lambda_n\alpha} (k,P) \Big[\frac{\partial h(p)}{\partial P_0}\widetilde{V}_{\lambda_n\lambda_{n}',\alpha\alpha'}(k,k';P)h(p')\Big]{\it \Psi}^{\lambda'}_{\alpha'\lambda_n'}(k',P)
\nonumber\\
&&\qquad\qquad\qquad\qquad\qquad\qquad\qquad\qquad\qquad+ \overline{\it\Psi}^{\lambda}_{\lambda_n'\alpha'} (k,P) \Big[h(p') \widetilde{V}_{\lambda_n'\lambda_{n},\alpha'\alpha}(k',k;P)\frac{\partial h(p)}{\partial P_0}\Big]{\it \Psi}^{\lambda'}_{\alpha\lambda_n}(k,P)\Big)\Big\}\, , \quad\label{eq:315first}
\eea
%
%
where use has been made of the fact that $p=P-k$ to replace $\partial/\partial p^0\to \partial/\partial P^0$.  In the second line we used the definition of ${\cal G}$ [recall Eq.~(\ref{eq:27})] and in the third line we replaced the reduced $\widetilde{\cal G}={\cal G}/h$
by the integral equation from which is is calculated.  Finally, renaming some repeated indices, and defining the derivative of the kernel with the reduced part held constant
\bea
\frac{\partial}{\partial P^0}\Big|_{\widetilde V}\overline{V}_{\lambda_n\lambda_{n}',\alpha\alpha'}(k,k';P)= \frac{\partial h(p)}{\partial P_0}\widetilde{V}_{\lambda_n\lambda_{n}',\alpha\alpha'}(k,k';P)\,h(p') + h(p)\, \widetilde{V}_{\lambda_n\lambda_{n}',\alpha\alpha'}(k,k';P) \frac{\partial h(p')}{\partial P_0} \label{eq:P0deriv}
\eea
gives
\bea
2e_0 \int_k &&\overline{\it\Psi}^{\lambda}_{\lambda_n\alpha} (k,P)\,j^0_{\alpha\alpha'}(p,p)\,{\it\Psi}^{\lambda'}_{\alpha'\lambda_n}(k,P)
\nonumber\\
=&&2e_0 \int_k \overline{\it\Psi}^{\lambda}_{\lambda_n\alpha} (k,P)\,\gamma^0_{\alpha\alpha'}\,{\it\Psi}^{\lambda'}_{\alpha'\lambda_n}(k,P) 
-2e_0\int_k\int_{k'} \overline{\it\Psi}^{\lambda}_{\lambda_n\alpha} (k,P) \frac{\partial}{\partial P^0}\Big|_{\widetilde V}\overline{V}_{\lambda_n\lambda_{n}',\alpha\alpha'}(k,k';P){\it \Psi}^{\lambda'}_{\alpha'\lambda_n'}(k',P)
\, . \label{eq:315firsta}
\eea
Using this result, and adding in the second term from Eq.~(\ref{eq:315}), which together with (\ref{eq:P0deriv}) gives the derivative of the full kernel $\overline V$, gives the alternate form for the normalization condition
\bea
2m_d\,e_d=2e_0 \int_k \overline{\it\Psi}^{\lambda}_{\lambda_n\alpha} (k,P)\,\gamma^0_{\alpha\alpha'}\,{\it\Psi}^{\lambda'}_{\alpha'\lambda_n}(k,P) 
-2e_0\int_k\int_{k'} \overline{\it\Psi}^{\lambda}_{\lambda_n\alpha} (k,P) \frac{\partial}{\partial P^0}\Big[\overline{V}_{\lambda_n\lambda_{n}',\alpha\alpha'}(k,k';P)\Big]{\it \Psi}^{\lambda'}_{\alpha'\lambda_n'}(k',P)
\, .\quad
\eea
The implications of the equivalence this form of the normalization condition with Eq.~(\ref{eq:315}) has already been discussed in Sec.~\ref{sec:charge}.

\end{widetext}

\section{Cancellations for terms depending only on momentum transfer} \label{app:intercur}

As stated in Sec.~\ref{sec:EX1} any kernel that depends only on the exchanged momentum will not contribute to the right-hand side of the two-body WT identity (\ref{eq:210}). To prove this statement, use (\ref{eq:4.1a}) to write the interaction as
\bea
\widetilde V_{\beta\beta',\alpha\alpha'}(k,p;k',p')&=&\frac12\Big[{V}_{\beta\beta',\alpha\alpha'}(k,p;k',p')
\nonumber\\&&
\qquad\pm{V}_{\alpha\beta',\beta\alpha'}(p,k;k',p')\Big]
\nonumber\\
&=&\frac12\Big[{V}_{\beta\beta',\alpha\alpha'}(q_d)\pm{V}_{\alpha\beta',\beta\alpha'}(q_e)\Big]\qquad
 \label{eq:4.1}
\eea
where the first term is the direct term (with momenta and Dirac indices labeled as in $\widetilde V$) and the second is the exchange term (with momenta and Dirac indicies of the final state particles exchanged from $\widetilde V$).  Because four-momentum is conserved in the CST, the momentum transfer for the direct and exchange terms can be written in two different ways
\bea
&&q_d=k'-k=p-p'
\nonumber\\
&&q_e=k'-p=k-p'.
\eea
Using this property we see immediately that the following relations hold (where the Dirac indices, the same for all terms, are suppressed, but the momentum labeling shows that the first two lines are for direct terms and the last two for exchange terms)
\bea
{V}(k,p_-;k',p'_-)&=&{V}(k,p_+;k',p'_+)={V}(k'-k)
\nonumber\\
{V}(k-q,p_+;k',p'_-)&=&{V}(k,p_+;k'+q,p'_-)
\nonumber\\&=&
{V}(k'-k+q)
\nonumber\\
{V}(p_-,k;k',p'_-)&=&{V}(p_+,k;k'+q,p'_-)
\nonumber\\&=&
{V}(k'+k-P_-)
\nonumber\\
{V}(p_+,k-q;k',p'_-)&=&{V}(p_+,k;k',p'_+)
\nonumber\\&=&
{V}(k'+k-P_+). \label{eq:qrelations}
\eea
Denoting the indices $\{\beta \beta',\alpha\alpha'\}$ by $d$ (for direct), and $\{\alpha\beta',\beta\alpha'\}$ by $e$ (for exchange), and using the relations (\ref{eq:qrelations}), the two body WT identity (\ref{eq:210}) can be written
\begin{widetext}
\begin{align}
q_\mu \widetilde{V}^\mu_{\beta\beta',\alpha\alpha'}(k,p_+;k',p'_-)
=&\frac{e_0}{2}\Big[{V}_d(k, p_-;k', p'_-)\pm V_e(p_-,k;k',p'_-)-{V}_d(k,p_+;k', p'_+)\mp V_e(p_+,k;k',p'_+)
\nonumber\\& 
+ {V}_d(k-q, p_+;k', p'_-)\pm V_e(p_+,k-q;k',p'_-)
 -{V}_d(k, p_+;k'+q, p'_-)\mp V_e(p_+,k;k'+q,p'_-)\Big]
 \nonumber\\
 =&\frac{e_0}{2}\Big[V_d(k'-k)\pm V_e(k'+k-P_-)-V_d(k'-k)\mp V_e(k'+k-P_+) 
\nonumber\\& 
+V_d(k'-k+q)\pm V_e(k'+k-P_+)-V_d(k'-k+q)\mp V_e(k'+k-P_-)\Big]=0,
\label{eq:210aa}
\end{align}
\end{widetext}
where, in the last expression, the four terms in the first line come from diagram (A) and the last four from diagrams (B)$_\pm$.  Both direct and exchange terms cancel in pairs, but the direct terms from diagrams (A) and (B) cancel separately, while the cancellation of the exchange terms requires contributions from both diagrams.  Note that this cancellation takes place {\it even though the  exchange terms depend on the total momemtum $P_+$ and $P_-$\/}.

\section{The $q\to0$ limit of Eq.~(\ref{eq:Icurrent})}\label{sec:qto0}

The $q\to0$ limit of (\ref{eq:Icurrent}) follows straightforwardly from the limit $j^\mu_0(0)=e_0\gamma^\mu$. Letting $P_+$ and $P_-$ approach $P$ gives
\begin{widetext}
\bea
\widetilde{V}^\mu_{\beta\beta',\alpha\alpha'}(kP; k' P)&=&\frac{e_0}{2m}\Big\{{V}^{2i}_{\beta\beta',\alpha\gamma}(k, k';P)\,\gamma^\mu_{\gamma\alpha'}+\gamma^\mu_{\alpha\gamma}\,{V}^{2f}_{\beta\beta',\gamma\alpha'}(k, k';P)
\nonumber\\&&
+{V}^{1i}_{\beta\gamma,\alpha\alpha'}(k, k';P)\,\gamma^\mu_{\gamma\beta'}+\gamma^\mu_{\beta\gamma}\,{V}^{1f}_{\gamma\beta',\alpha\alpha'}(k, k';P) \Big\}\, .\label{eq:Icurrentat0}
\eea
To demonstrate that this is equivalent to (\ref{eq:211}), consider the derivatives of the reduced kernel.  Introducing the operator 
\bea
D^\mu=-2e_0\frac{\partial}{\partial P_\mu}-e_0\frac{\partial}{\partial k_\mu}-e_0\frac{\partial}{\partial k'_\mu}\, .
\eea
and using the fact that the action of this operator on any terms that depend only on the momentum  of the exchanged meson (for example $(k'-k)^2$ for the direct terms or $(P-k-k')^2$ for exchange terms) will give zero, means that only the terms with a factor of $\Theta$ will contribute.  Furthermore, using the fact that 
\bea
D^\mu \Theta(k)=D^\mu \Theta(k')=D^\mu \Theta(P-k)=D^\mu \Theta(P-k')=e_0\frac{\gamma^\mu}{2m}
\eea
permits the result to be written directly in terms of the truncated kernels.  If $k$ and $k'$ are off-shell, so that the kernel $\widetilde V$ includes the factors of $\Theta(k)$ and $\Theta(k')$, then the result is
\bea
D^\mu\widetilde{V}_{\beta\beta',\alpha\alpha'}(k, k';P)\Big|_{k,k' {\rm off}}&=&\frac{e_0}{2m}\Big\{{V}^{2i}_{\beta\beta',\alpha\gamma}(k, k';P)\,\gamma^\mu_{\gamma\alpha'}+\gamma^\mu_{\alpha\gamma}\,{V}^{2f}_{\beta\beta',\gamma\alpha'}(k, k';P)
\nonumber\\&&
+{V}^{1i}_{\beta\gamma,\alpha\alpha'}(k, k';P)\,\gamma^\mu_{\gamma\beta'}+\gamma^\mu_{\beta\gamma}\,{V}^{1f}_{\gamma\beta',\alpha\alpha'}(k, k';P) \Big\}\, ,\label{eq:Icurrentat1}
\eea
proving that (\ref{eq:Icurrent}) and (\ref{eq:211}) are identical.  {\it However\/}, if particle 1 is on-shell, the kernel will not include any factors of $\Theta(k)$ and $\Theta(k')$, so that the result is
\bea
D^\mu\widetilde{V}_{\beta\beta',\alpha\alpha'}(k, k';P)\Big|_{k,k' {\rm on}}&=&\frac{e_0}{2m}\Big\{{V}^{2i}_{\beta\beta',\alpha\gamma}(k, k';P)\,\gamma^\mu_{\gamma\alpha'}+\gamma^\mu_{\alpha\gamma}\,{V}^{2f}_{\beta\beta',\gamma\alpha'}(k, k';P)
\Big\}\, .\label{eq:Icurrentat2}
\eea
%


\end{widetext}

\end{document}